\documentclass[iop]{emulateapj}
\usepackage{apjfonts}
\usepackage{graphicx}
\usepackage{hyperref}

\def\apj {ApJ}
\def\apjl {ApJL}
\def\apjs {ApJS}
\def\aj {AJ}
\def\mnras {MNRAS}
\def\aap {A\&A}
\def\nat {Nat}

\def\araa {ARAA}

\def\pasj {PASJ}

\def\bea		{\begin{eqnarray}}
\def\eea		{\end{eqnarray}}
\def\bee		{\begin{equation}}
\def\eee		{\end{equation}}
\def\bef		{\begin{figure}}
\def\eef		{\end{figure}}
\def\befs	{\begin{figure*}}
\def\eefs	{\end{figure*}}
\def\be		{\begin{equation}}
\def\ee		{\end{equation}}

\def\CII  {\ifmmode{{\rm C\:\scriptstyle II}}\else{${\rm C\:\scriptstyle II}$}\fi}
\def\CIII  {${\rm C\:\scriptstyle III}$}
\def\CIV {\ifmmode{{\rm C\:\scriptstyle IV}}\else{${\rm C\:\scriptstyle IV}$}\fi}
\def\CV {\ifmmode{{\rm C\:\scriptstyle V}}\else{${\rm C\:\scriptstyle V}$}\fi}
\def\SiII  {${\rm Si\:\scriptstyle II}$}
\def\SiIII  {${\rm Si\:\scriptstyle III}$}
\def\SiIV  {${\rm Si\:\scriptstyle IV}$}

\def\OVI {${\rm O\:\scriptstyle VI}$}
\def\OVII {${\rm O\:\scriptstyle VII}$}
\def\OVIII {${\rm O\:\scriptstyle VIII}$}
\def\HI   {${\rm H\:\scriptstyle\rm I}$}
\def\HII   {${\rm H}^+$}

\def\ff   {\ifmmode{f}\else{$f$}\fi}

\def\rmax {\ifmmode{r_{\rm max}}\else{$r_{\rm max}$}\fi}
\def\zmax {\ifmmode{z_{\rm max}}\else{$z_{\rm max}$}\fi}

\def\fCNM {\ifmmode{f^{\rm CNM}}\else{$f^{\rm CNM}$}\fi}
\def\fCMM {\ifmmode{f^{\rm CMM}}\else{$f^{\rm CMM}$}\fi}
\def\fWNM {\ifmmode{f^{\rm WNM}}\else{$f^{\rm WNM}$}\fi}
\def\fWIM {\ifmmode{f^{\rm WIM}}\else{$f^{\rm WIM}$}\fi}
\def\fCM  {\ifmmode{f^{\rm CM}} \else{$f^{\rm CM}$}\fi}
\def\fWM  {\ifmmode{f^{\rm WM}} \else{$f^{\rm WM}$}\fi}

\def\noCNM {\ifmmode{n_o^{\rm CNM}}\else{$n_o^{\rm CNM}$}\fi}
\def\noCMM {\ifmmode{n_o^{\rm CMM}}\else{$n_o^{\rm CMM}$}\fi}
\def\noWNM {\ifmmode{n_o^{\rm WNM}}\else{$n_o^{\rm WNM}$}\fi}
\def\noWIM {\ifmmode{n_o^{\rm WIM}}\else{$n_o^{\rm WIM}$}\fi}
\def\noCM  {\ifmmode{n_o^{\rm CM}} \else{$n_o^{\rm CM}$}\fi}
\def\noWM  {\ifmmode{n_o^{\rm WM}} \else{$n_o^{\rm WM}$}\fi}

\def\nnCNM {\ifmmode{n^{\rm CNM}}\else{$n^{\rm CNM}$}\fi}
\def\nnCMM {\ifmmode{n^{\rm CMM}}\else{$n^{\rm CMM}$}\fi}
\def\nnWNM {\ifmmode{n^{\rm WNM}}\else{$n^{\rm WNM}$}\fi}
\def\nnWIM {\ifmmode{n^{\rm WIM}}\else{$n^{\rm WIM}$}\fi}
\def\nnCM  {\ifmmode{n^{\rm CM}} \else{$n^{\rm CM}$}\fi}
\def\nnWM  {\ifmmode{n^{\rm WM}} \else{$n^{\rm WM}$}\fi}

\def\zCNM {\ifmmode{z_o^{\rm CNM}}\else{$z_o^{\rm CNM}$}\fi}
\def\zCMM {\ifmmode{z_o^{\rm CMM}}\else{$z_o^{\rm CMM}$}\fi}
\def\zWNM {\ifmmode{z_o^{\rm WNM}}\else{$z_o^{\rm WNM}$}\fi}
\def\zWIM {\ifmmode{z_o^{\rm WIM}}\else{$z_o^{\rm WIM}$}\fi}
\def\zCM  {\ifmmode{z_o^{\rm CM}} \else{$z_o^{\rm CM}$}\fi}
\def\zWM  {\ifmmode{z_o^{\rm WM}} \else{$z_o^{\rm WM}$}\fi}

\def\nH 		{\ifmmode{\langle n_{\rm H} \rangle}\else{$\langle n_{\rm H} \rangle$}\fi}
\def\ne 		{\ifmmode{\langle n_{\rm e} \rangle}\else{$\langle n_{\rm e} \rangle$}\fi}
\def\Em 		{\ifmmode{E_m}\else{$E_m$}\fi}
\def\deg   	{\ifmmode{^\circ}\else{$^\circ$}\fi}
\def\fE    	{\ifmmode{f_{\rm E}}\else{$f_{\rm E}$}\fi}
\def\Trec 	{\ifmmode{T_{\rm rec}}\else{$T_{\rm rec}$}\fi}
\def\Ncr    	{\ifmmode{N_{\rm cr}}\else{$N_{\rm cr}$}\fi}
\def\Np    	{\ifmmode{N_{\rm p}}\else{$N_{\rm p}$}\fi}
\def\HH    	{\ifmmode{\cal H}\else{${\cal H}$}\fi}
\def\dH    	{\ifmmode{\delta\cal H}\else{$\delta{\cal H}$}\fi}
\def\Hdisk 	{\ifmmode{\cal H}_{\rm disk}\else{${\cal H}_{\rm disk}$}\fi}
\def\Hdiskp {\ifmmode{\cal H}^{'}_{\rm disk}\else{${\cal H}^{'}_{\rm disk}$}\fi}
\def\Hsky  	{\ifmmode{\cal H}_{\rm sky}\else{${\cal H}_{\rm sky}$}\fi}

\def\gta{\;\lower 0.5ex\hbox{$\buildrel > \over \sim\ $}}
\def\lta{\;\lower 0.5ex\hbox{$\buildrel < \over \sim\ $}}          
\def \NH {\ifmmode{\rm N}_{\scriptscriptstyle H}\else{N$_{\scriptscriptstyle H}$}\fi}

\def\Msun   {\ M$_{\odot}$}

\def\kms    {\ km s$^{-1}$}

\newcommand	{\Ha}{\mbox {H$\alpha$}}

\def\deg		{\hbox{${}^\circ$}}

\def\las		{\mathrel{\hbox{\rlap{\hbox{\lower4pt
        \hbox{$\sim$}}}\hbox{$<$}}}}
\def\gas		{\mathrel{\hbox{\rlap{\hbox{\lower4pt
            \hbox{$\sim$}}}\hbox{$>$}}}}

\begin{document}

\title{The large-scale ionization cones in the Galaxy}

\author{}
\affil{}
\author{Joss Bland-Hawthorn\altaffilmark{1,2}}
\affil{Sydney Institute for Astronomy, School of Physics A28,
  University of Sydney, NSW 2006, Australia  \\
  $^1$Miller Professor, Miller Institute, University of California, Berkeley, CA 94720, USA \\
  $^2$ARC Centre of Excellence for All Sky Astrophysics in 3 Dimensions (ASTRO-3D), Australia} 
  \author{Philip R. Maloney} 
\affil{Boulder, CO 80301, USA}\author{Ralph Sutherland\altaffilmark{3} \& Brent Groves}
\affil{Research School of Astronomy \& Astrophysics, Australia National University, Canberra 2611, Australia \\
    $^3$ARC Centre of Excellence for All Sky Astrophysics in 3 Dimensions (ASTRO-3D), Australia}
  \author{Magda Guglielmo, Wen Hao Li \& Andrew Curzons}
\affil{Sydney Institute for Astronomy, School of Physics A28,
  University of Sydney, NSW 2006, Australia} 
\author{Gerald Cecil}
\affil{Department of Physics and Astronomy,
University of North Carolina, Chapel Hill, NC, USA}
  \author{Andrew J. Fox}
\affil{Space Telescope Science Institute, 3700 San Martin Drive, Baltimore, MD 21218, USA}

\begin{abstract}
\vspace{0.1in} There is compelling evidence for a highly energetic Seyfert explosion ($10^{56-57}$ erg) that occurred in the Galactic Centre a few million years ago. The clearest indications are the x-ray/$\gamma$-ray `10 kpc bubbles' identified by the {\it Rosat} and {\it Fermi} satellites. In an earlier paper, we suggested another manifestation of this nuclear activity, i.e. elevated \Ha\ emission along a section of the Magellanic Stream due to a burst (or flare) of ionizing radiation from Sgr A*. We now provide further evidence for a powerful flare event: UV absorption
line ratios (in particular \CIV/\CII, \SiIV/\SiII) observed by the {\it Hubble Space Telescope} reveal that some Stream clouds towards both galactic poles are highly ionized by a source capable of producing ionization energies up to at least 50 eV. We show how these are clouds caught in a beam of bipolar, radiative `ionization cones' from a Seyfert nucleus associated with Sgr A*. In our model,
the biconic axis is tilted by about 15\deg\ from the South Galactic Pole
with an opening angle of roughly 60\deg. For the Stream at such large Galactic distances ($D \gta 75$ kpc),
nuclear activity is a plausible explanation for all of the
observed signatures: elevated \Ha\ emission and H ionization fraction ($x_e \gta 0.5$), enhanced \CIV/\CII\ and \SiIV/\SiII\ ratios, and high \CIV\ and \SiIV\ column densities.
Wind-driven `shock cones' are ruled out because the \textit{Fermi} bubbles lose their momentum and energy to the Galactic corona long before reaching the Stream.
Our time-dependent Galactic ionization model (stellar populations, hot coronal gas, cloud-halo interaction) is too weak to explain the Stream's ionization. Instead, the nuclear flare event must have had a radiative UV luminosity close to the Eddington limit ($f_E \approx 0.1-1$). 
Our time-dependent Seyfert flare models adequately explain the observations and indicate the Seyfert flare event took place $T_o = 3.5\pm 1$ Myr ago. The timing estimates are consistent with the mechanical timescales needed to explain the x-ray/$\gamma$-ray bubbles in leptonic jet/wind models ($\approx 2-8$ Myr).
\end{abstract}

\begin{figure*}
\centering
\includegraphics[scale=0.15]{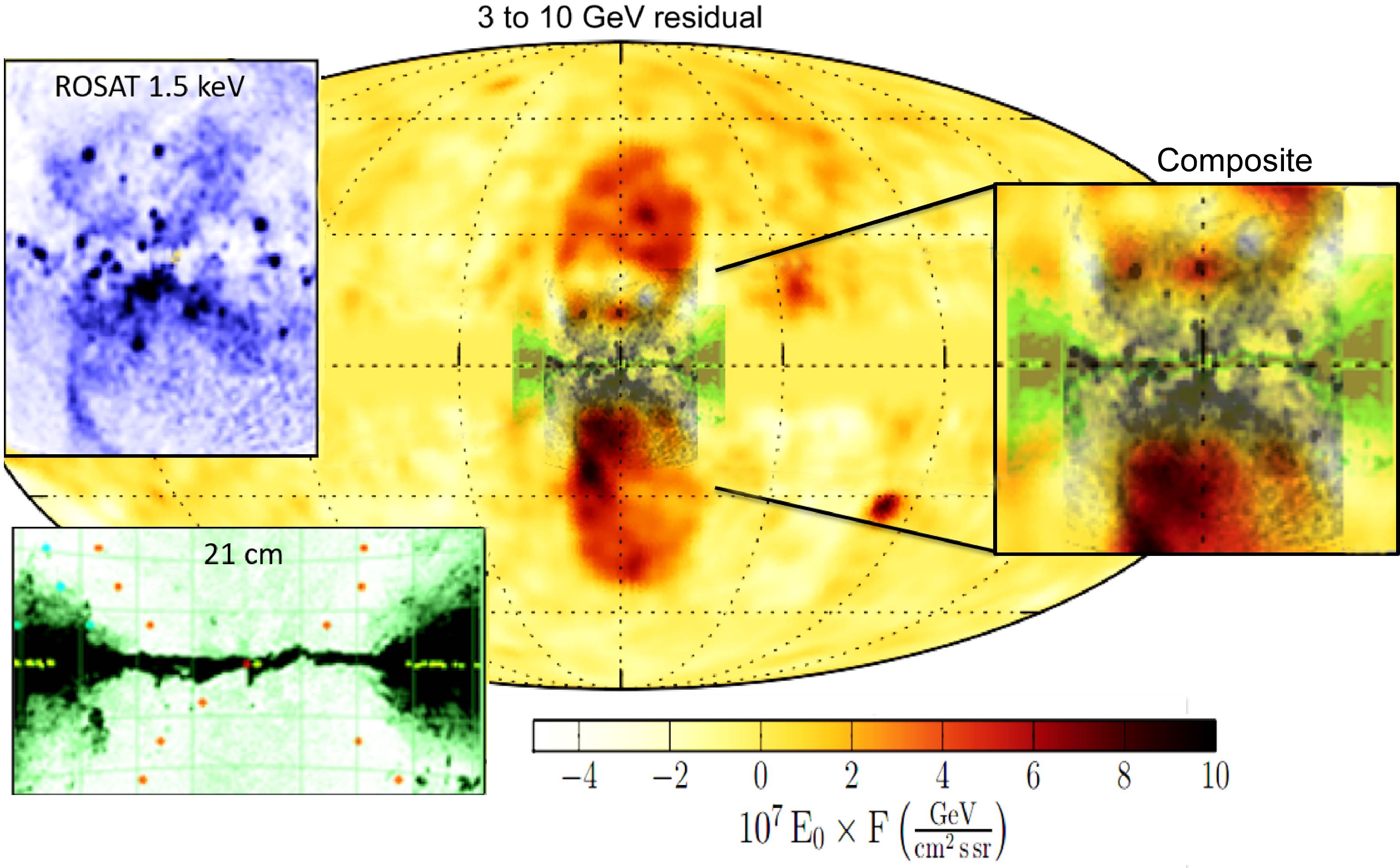}
\caption{All-sky Mollweide projection (NGP uppermost) aligned with the Galactic Centre showing the strong association between
the 3-10 GeV $\gamma$-ray emission (Ackermann et al 2014 $-$ main image), the 1.5 keV
x-ray emission (Bland-Hawthorn \& Cohen 2003 $-$ blue inset), and
the 21cm cold hydrogen emission (Lockman \& 
McClure-Griffiths 2016 $-$ green inset with orange dots spaced 1 kpc apart at the distance of the
Galactic Centre). On the RHS, 
we show a magnified region  around the Galactic Centre as a colour composite with
all three components overlaid. 
}
\label{f:fermi}
\end{figure*}

\begin{figure*}
\centering
\includegraphics[scale=0.7]{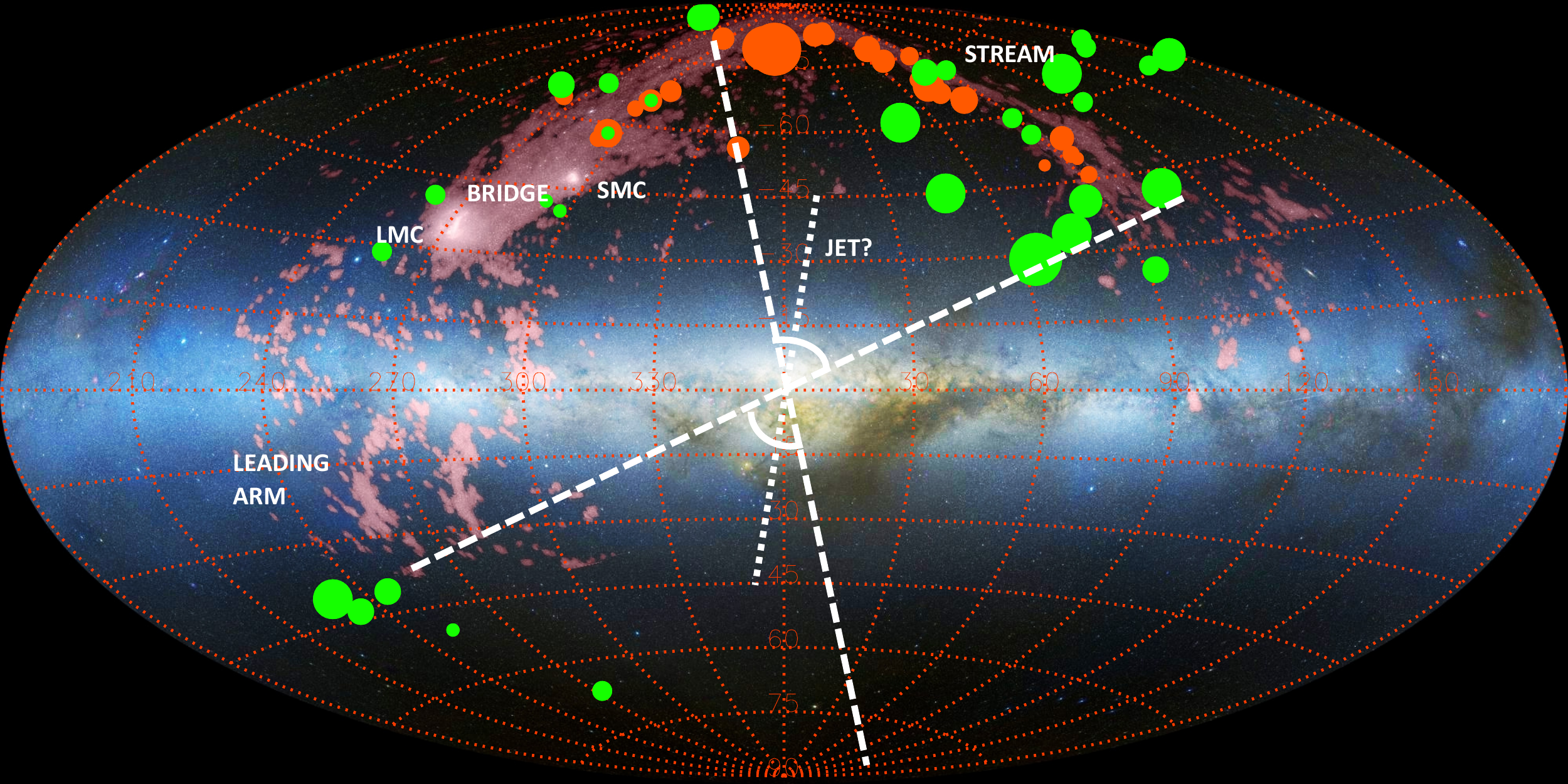}
\caption{Rotated all-sky Aitoff projection (South Galactic Pole uppermost) aligned with the 
Galactic Centre showing the orientation of
the ionization cones (\S 4) inferred from this work. The 3D space orientation is 
uncertain: the opening angle is roughly 60$^\circ$
and includes the Galactic polar axis. The red points indicate the \Ha\ detections where the symbol size scales with the surface brightness; the green points scale with the strength of the \CIV/\CII\ ratio with larger points indicating a harder radiation field (if photoionized).
The optical image and 21cm overlay (pink) was first presented by Nidever et al (2008); the radio emission is from the 21cm \HI\ mapping of
the Magellanic Clouds and Stream (including the
leading arms) by Kalberla et al (2005). Note that some Stream \HI\ clouds fall within the cones (indicated by small arcs) in both hemispheres. The dotted line indicates the axis of a putative radio/$\gamma$-ray jet (Bower \& Backer 1998; Su \& Finkbeiner 2012).}
\label{f:nidever}
\end{figure*}

\section{Introduction}
One of the oldest baryonic remnants of the early universe in our Galaxy is the massive black hole in Sgr A*. At redshifts higher than $z\sim 4$, black holes are thought to grow rapidly through radiatively inefficient accretion (Inayoshi et al 2016) and the merger of subsystems harbouring lower mass black holes
(Volonteri 2010).
After that time, the growth is regulated by the infall of gas, stars and dark matter. The last few e-folds of mass over 10 Gyr are grown via radiatively {\it efficient} accretion (Rees \& Volonteri 2007). The conversion efficiency must be $\epsilon\approx 10\%$ to explain the UV/x-ray background (Soltan 1982; Yu \& Tremaine 2002; Zhang \& Lu 2019). A black hole with mass $M_\bullet$ today has converted $\epsilon M_\bullet c^2$ of its rest mass into emergent energy.
Over the past 10 Gyr, Sgr A*, for which $M_{\bullet} = 4.15\times 10^6$\Msun\ (Gravity Collaboration 2019), must have released $\sim 10^{60}$ erg in relativistic particles and electromagnetic radiation to get to its current state.

In the Milky Way, we observe the x-ray/$\gamma$-ray bubbles with an inferred energy 10$^{56-57}$ erg. The first evidence of a kiloparsec-scale outflow in the Galaxy came from bipolar {\it Rosat} 1.5 keV x-ray emission inferred to be associated with the Galactic Centre (Bland-Hawthorn \& Cohen 2003). In Fig.~\ref{f:fermi}, this same component is 
directly associated with the {\it Fermi} $\gamma$-ray bubbles (1-100 GeV) discovered by Su et al (2010). Star formation activity fails on energetic grounds by a factor of 400 based on what we see today (Miller \& Bregman 2016), or $\sim$100 if we allow for past starbursts within the limits imposed by the resolved stellar population (Nataf 2016; Bland-Hawthorn et al 2013, hereafter BH2013).

The source of the x-ray/$\gamma$-ray bubbles can only be from nuclear activity: all contemporary leptonic models of the x-ray/$\gamma$-ray bubbles agree on this point with timescales for the event falling in the range 2 to 8 Myr (Guo \& Matthews 2012; Miller \& Bregman 2016; Narayanan \& Slatyer 2017; cf. Carretti et al 2012). 
These must be driven by the AGN (jet and/or accretion-disk wind) on a timescale of order a few Myr $-$ for a comprehensive review, see Yang et al (2018). 

AGN jets are remarkably effective at blowing bubbles regardless of the jet orientation because the jet head is diffused or deflected by each interaction with density anomalies in a fractal ISM (Zovaro et al 2019). The evidence for an active jet today in the Galactic Centre is weak (Bower \& Backer 1998). Su \& Finkbeiner (2012) found a jet-like feature in $\gamma$-rays extending from $({\ell,b})$ $\approx$ (-11\deg, 44\deg) to (11\deg, -44\deg); this axis
is indicated in Fig.~\ref{f:nidever}. In recent simulations, the AGN jet drills its way through the multiphase ISM with a speed of roughly 1 kpc per Myr (Mukherjee et al 2018, Appendix A). 
If the tentative claims are not confirmed, this
may indicate that either the AGN outflow was not accompanied by a jet, or the jet has already pushed through the inner disk gas and has now dispersed.

Absorption-line UV spectroscopy of background AGN and halo stars reveals cool gas clouds entrained in the outflow (Fox et al 2015; Bordoloi et al 2017; Savage et al 2017; Karim et al 2018); \HI\ clouds are also seen (Di Teodoro et al 2018). Modeling of the cloud kinematics yields similar timescales for the wind 
($\sim$6--9 Myr; Fox et al. 2015, Bordoloi et al. 2017). An updated shock model of the \OVII\ and \OVIII\ x-ray emission over the bubble surfaces indicates the initial burst took place $4\pm 1$ Myr ago (Miller \& Bregman 2016).

A number of authors (e.g. Zubovas \& Nayakshin 2012) tie the {\it localized} x-ray/$\gamma$-ray activity to the formation of the young stellar annulus ($M_\star \sim 10^4$\Msun) in orbit about the Galactic Centre (q.v. Koyama 2018). These stars, with uncertain ages in the range 3-8 Myr, are mostly on elliptic orbits and stand out in a region dominated by an old stellar population (Paumard et al 2006; Yelda et al 2014). A useful narrative of how this situation can arise is given by Lucas et al (2013): a clumpy prolate cloud with a dimension of order the impact radius, and oriented perpendicular to the accretion plane, sets up accretion timescales that can give rise to high-mass stars in elliptic prograde and retrograde orbits.

Nuclear activity peaked during the golden age of galaxy formation ($z=1-3$; Hopkins \& Beacom 2006), but it is observed to occur in a few percent of galaxies at lower levels today. Given that most galaxies possess nuclear black holes, this activity may be ongoing and stochastic in a significant fraction, even if only detectable for a small percentage of sources at a given epoch (Novak et al 2011). If most of the activity occurred after $z \sim 1$, this argues for \textit{Fermi} bubble-like outbursts roughly every $\sim$10 Myr or so. Each burst may have lasted up to $\sim$1 Myr at a time (Guo \& Mathews 2012), flickering on shorter timescales. This argues that $\sim$10\% of all galaxies are undergoing a Seyfert phase at any time but where most outcomes, like the \textit{Fermi} bubbles, are not easily detectable (cf. Sebastian et al 2019).

Independent of the mechanical timescales, BH2013 show that the high levels of \Ha\ emission along the Magellanic Stream are consistent with a Seyfert ionizing flare event 2-3 Myr ago (see Fig.~\ref{f:nidever}); starburst-driven radiation fails by two orders of magnitude. Ionisation cones are not uncommon in active galaxies today (e.g. Pogge 1988; Tsvetanov et al 1996) and can extend to $\sim$100 kpc distances (Kreimeyer \& Veilleux 2013). Here we revisit our earlier work in light of new observations and a better understanding of the Magellanic Stream's distance from the Galaxy.
In \S 2, we update what has been learnt about the ionization, metallicity and gas content of the Magellanic Stream and its orbit properties. 
\S 3 builds up a complete model of the 
Galactic UV radiation field and include a major
AGN contribution to illustrate the impact of nuclear activity. In \S 4,
we carry out time-dependent ionization calculations
to update the likely timescale for the Seyfert flare. \S 5 concludes with suggested follow-up
observations and discusses the implications of our findings.

\section{New observations}

\subsection{Magellanic Stream: gas and metal content}

Since its discovery in the 1970s, many authors have studied the physical
properties of the gas along the Magellanic Stream (Putman et al 1998;
Br\"{u}ns et al 2005; Kalberla et al 2005; Stanimirovic et al 2008; Fox et al 2010; Nigra et al 2012). The Stream lies along a great arc that
spans more than half the sky (e.g. Nidever et al 2010). Its metallicity content is generally about one 
tenth of the solar value, consistent with the idea that the gas came
from the SMC and/or the outer regions of the LMC
(Fox et al 2013), although a filament tracing back to the inner LMC has an elevated level of metal enrichment (Richter et al 2013).
The inferred total mass of the Magellanic Stream is ultimately linked to its distance $D$ from the Galactic Centre. The total \HI\ mass of the Magellanic gas system
(corrected for He) is $5\times 10^8\; (D/55\; {\rm kpc})^2$\Msun\ 
(Br\"{u}ns et al 2005) but this may not even
be the dominant mass component (Bland-Hawthorn et al 2007;
d'Onghia \& Fox 2016). Fox et al (2014) find that the
plasma content may dominate over the neutral gas by a factor
of a few such that the Stream's total gas mass is closer 
to $2.0\times 10^9\; (D/55\; {\rm kpc})^2$\Msun. We discuss the likely
value of $D$ measured along the South Galactic Pole (SGP) in the next section.

\subsection{Magellanic Stream: orbit trajectory}

The precise origin of the trailing and leading arms of the Magellanic 
Stream is unclear. Theoretical models for the Stream date back
to at least the early seminal work of Fujimoto \& Sofue (1976, 1977).
For three decades, in the absence of a distance indicator, 
the Stream's distance over the SGP was traditionally 
taken to be the midpoint in the LMC and SMC distances, i.e. $D=55$ kpc,
a distance which is now thought to be too small.

In a series of papers, Kallivayalil and collaborators show that the proper motions of the LMC and SMC are 30\% higher than original longstanding estimates (e.g. Kallivayalil et al 2006, 2013). Over the same period, mass estimates of the Galaxy have decreased to $M_{\rm vir} = 1.3\pm 0.3\times 10^{12}$\Msun\ (q.v. Bland-Hawthorn \& Gerhard 2016; McMillan 2017). Thus the orbit of the Magellanic System must be highly elliptic. Contemporary models consider the LMC and SMC to be on their first infall with an orbital period of order a Hubble time (Besla et al 2007, 2012; Nichols et al 2011). 

The Stream is a consequence of the tidal interaction between both dwarfs. The models move most of the trailing Stream material to beyond 75 kpc over the SGP. Here we take a representative model for the Stream particles from a recent hydrodynamical simulation (Guglielmo et al 2014), adopting a smooth fit to the centroid of the particle trajectory and some uncertainty about that trajectory. 

In passing, we note that while the trailing Stream is understood in these models, the `leading arm' is unlikely to be explaine as a tidal extension in the same way because of the strong ram-pressure confinement imposed by the Galactic corona ahead of the Magellanic Clouds (Tepper-Garcia et al 2019). It can be debris arising from an earlier interaction of the LMC-SMC system protected by a Magellanic corona, for example. Thus, the origin of the `leading arm' is unclear and its distance is poorly constrained. Most of the cool gas ahead of the Clouds lies outside of the ionization cones in Fig.~\ref{f:nidever}.

\subsection{Magellanic Stream: ionization}

Weiner \& Williams (1996) first discovered elevated levels of H$\alpha$ emission along the Magellanic Stream,
detections that have been confirmed and extended through follow-up observations (Weiner et al 2002; Putman et al 2003; BH2013; Barger et al 2017). There have been several attempts to understand this emission over the past two decades in terms of Galactic sources (Bland-Hawthorn \& Maloney 1999), particle trapping by magnetic fields (Konz et al 2001),
thermal conduction and mixing (Fox et al 2005), cloud-halo (Weiner \& Williams 1996) and cloud-cloud interactions (Bland-Hawthorn et al 2007).

While these sources can contribute to ionization and heating along the Magellanic Stream, in light of new evidence, we believe that only the Seyfert flare model (BH2013) survives as a likely candidate for the brightest emission. Further evidence for non-thermal photons being the source of the ionization comes from a
UV spectroscopic study carried out with the {\it Hubble Space Telescope} ({\it HST}) of distant quasars that lie behind the Magellanic Stream. Fox et al (2014) infer ionization levels along the Stream from UV
absorption features arising from \HI, \SiII, \SiIII, \SiIV, \CII\ and \CIV. They find that there are three patches along the Stream that require enhanced levels of hard ionization (30-50 eV photons) relative to stellar photons. One is highly localized at the LMC; the other regions lie towards the NGP and SGP. We argue below
that these regions fall within the `ionization cones' of the Seyfert event. These data are presented and modelled in \S 4.  

\begin{figure*}[hbtp]
\centering
\includegraphics[scale=0.4]{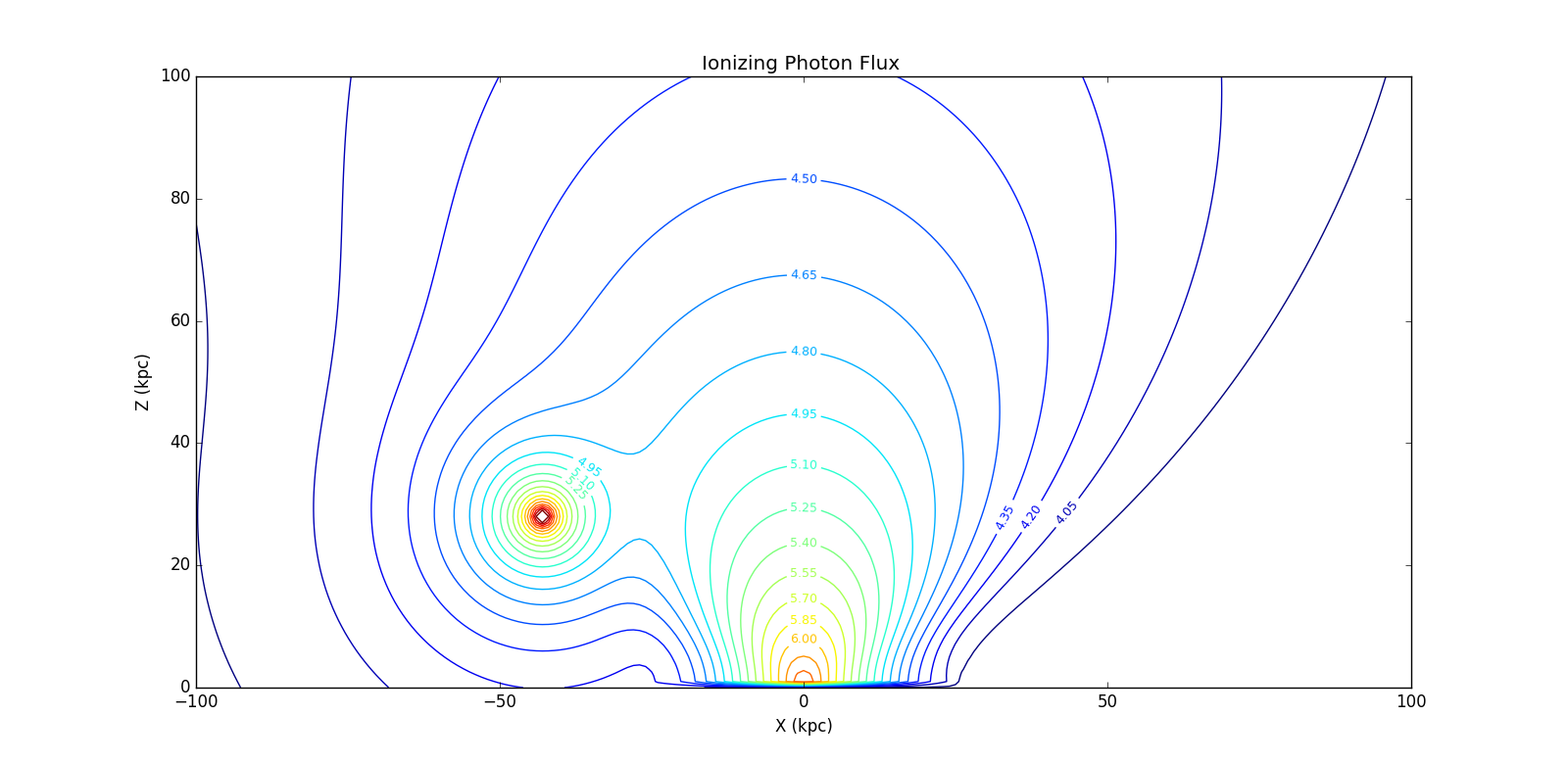}
\caption{Our model for the ionizing radiation field over
the South Galactic Hemisphere arising from the opaque Galactic disk and the Large Magellanic cloud (\S 3). The units of the contours are $\log$(phot cm$^{-2}$ s$^{-1}$). Small contributions from the hot Galactic corona
and the cosmic UV background are also included. The $X-Z$
plane runs through the LMC, the SGP and the Galactic Centre
defined by the plane of Magellanic longitude.
The ionizing flux contours are spaced in equal log intervals.}
\label{f:Galaxy}
\end{figure*}

\begin{figure*}[hbtp]
\centering
\includegraphics[scale=0.5]{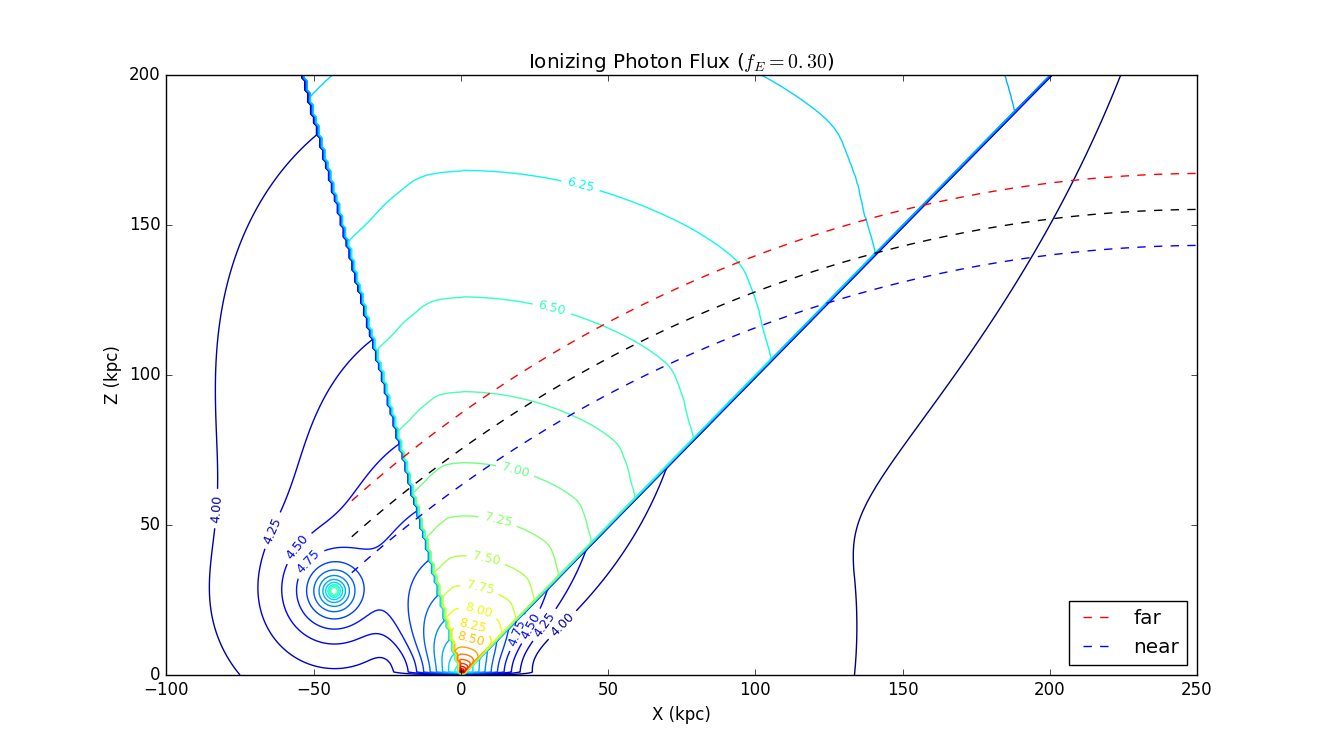}
\caption{The ionizing field presented in Fig.~\ref{f:Galaxy}
with the added contribution of a Seyfert flare event, but
shown on a larger physical scale. The units of the contours are $\log$(phot cm$^{-2}$ s$^{-1}$). 
For illustration, we show
the impact of a sub-Eddington flare ($f_E=0.3$). This flux level is needed to reproduce
what we observe but is inconsistent with Sgr A*
activity today. A more likely scenario is that the event 
occurred in the past and what is seen today is the fading
recombination of this flare (BH2013).
The black trajectory is a fit to the orbit path of the
Magellanic Stream particles (Guglielmo et al 2014) that uses updated parameters for the Galaxy and is quite typical
of modern simulations. The blue and red tracks represent
the $3\sigma$ uncertainties for the distribution of Stream particles. The ionizing flux contours are spaced in equal log intervals. A schematic movie of pulsing AGN radiation is available at
\url{http://www.physics.usyd.edu.au/~jbh/share/Movies/SgrA*_ionized_cone.gif}.
We also include a simulation of flickering AGN radiation (Novak et al 2011) impinging on the Magellanic Stream at
\url{http://www.physics.usyd.edu.au/~jbh/share/Movies/MilkyWay_ionized_cone.mp4}; the movie ends when the Magellanic Clouds reach their observed position today.
}
\label{f:Seyfert}
\end{figure*}

\section{New models}

\subsection{Galactic ionization model}

We model the Magellanic Stream H$\alpha$ emission and
carbon absorption features using the Galactic
ionization model presented by Bland-Hawthorn \& Maloney (1999, 2002), updated with the time-dependent calculations in BH2013.
A cross-section through the 3D model across the
South Galactic Hemisphere passing through the
Galactic Centre and the LMC
is shown in Fig.~\ref{f:Galaxy}.
The Galactic disk parameters remain unchanged from earlier work where we considered the expected emission arising from stars.
The total flux at a frequency $\nu$ reaching an observer
located at a distance $D$ is obtained from integrating the specific
intensity $I_\nu$ over the surface of a disk, i.e.
\bee
F_\nu = \int_A I_\nu({\bf n})({\bf n}.{\bf N}) {{dA}\over{D^2}}
\label{e:flux}
\eee
where ${\bf n}$ and ${\bf N}$ are the directions of the line of sight
and the outward normal to the surface of the disk, respectively.
In order to convert readily to an \Ha\ surface brightness, we 
transform equation~(\ref{e:flux}) to a photon flux   after including
the effect of disk opacity $\tau_{D}$ at the Lyman limit such that
\bee
\varphi_\star = \int_\nu {{F_\nu}\over{h\nu}} \exp(-\tau_{D}/\cos\theta)\ \cos \theta\ d\nu
\label{e:star}
\eee
for which $\vert \theta \vert \neq \pi/2$ and
where $\varphi_\star$ is the photoionizing flux from the stellar
population, ${\bf n}.{\bf N} = \cos\theta$ and $h$ is Planck's
constant. This is integrated over frequency above the Lyman limit
($\nu=13.6\; {\rm eV}/h$) to infinity to convert to units of photon
flux (phot cm$^{-2}$ s$^{-1}$). 
The mean vertical opacity of the disk over the
stellar spectrum is
$\tau_{D}=2.8\pm 0.4$, equivalent to a vertical escape fraction of
$f_{\star,{\rm esc}} \approx 6$\% perpendicular to the disk (${\bf
n}.{\bf N} = 1$).  

The photon spectrum of the Galaxy is a
complex time-averaged function of energy $N_\star$ (photon rate per
unit energy) such that $4\pi D^2 \varphi_\star = \int_0^{\infty}
N_\star(E)\: dE$.
For a given ionizing luminosity, we can determine the expected
H$\alpha$ surface brightness at the distance of the Magellanic Stream.
For an optically thick cloud ionized on one side, we relate the
emission measure to the ionizing photon flux using ${\cal E}_m =
1.25\times 10^{-6} \varphi_\star$ cm$^{-6}$ pc (Bland-Hawthorn \&
Maloney 1999). In Appendix A, we relate ${\cal E}_m$ 
to the more familiar milliRayleigh units (mR) used widely
in diffuse detection work.

The Galactic UV contribution at the distance $D$ of
the Magellanic Stream is given by
\bee
\mu_{\star,{\rm H}\alpha} = 10\zeta \left({f_{\star,{\rm
      esc}}}\over{0.06} \right) \left({D}\over{{\rm 75\
    kpc}}\right)^{-2} \ \ \ {\rm mR} . 
\label{f:Gal}
\eee
The correction factor $\zeta \approx 2$ is included to accommodate weakly
constrained ionizing contributions from old stellar populations and
fading supernova bubbles in the disk (BH2013).

After Barger et al (2013) and Fox et al (2014), we incorporate the UV contribution from the 
Large Magellanic Cloud (LMC) but with an important modification. 
Barger et al (2013) showed how the LMC UV ionizing intensity is
sufficient to ionize the Magellanic Bridge in close proximity; the SMC UV radiation field can
be neglected. This is 
assisted by the orientation of the LMC disk with respect to the Bridge. In our treatment, the LMC's greater distance and orientation does not assist the ionization of the Magellanic
Stream. We treat the LMC as a point source with a total ionizing luminosity 
reduced by a factor $\exp({-\tau_L}$); $\tau_L=1.7$ is the mean LMC disk
opacity which we scale from the Galactic disk opacity ($\tau_D=2.8$) by the
ratio of their metallicities (Fox et al 2014).

We stress that the LMC cannot be the source of the Magellanic Stream ionization. Its imprint over the local \HI\ is clearly seen in Barger et al (2013, Fig. 16). One interesting prospect, suggested by the referee, is that one or more ultraluminous x-ray sources (ULX) in the LMC have produced a flash of hard UV/x-ray radiation in the recent past. In fact, a few such sources are observed there (Kaaret et al 2017) and may be responsible for the localized \CIV/\CII\ enhancement at the LMC (Fox et al 2014). We include the super-Eddington accretion spectrum in our later models (\S 4.3, \S 4.4.1) to emphasize this point.

\smallskip
\noindent
{\sl Other sources.} We have used updated parameters for the Galactic 
corona from Miller \& Bregman (2016), but the
UV emission from the halo remains negligible (i.e. a few percent at most)
compared to the Galactic disk ($\varphi_\star \sim 5\times 10^4$ phot 
cm$^{-2}$ s$^{-1}$ at 75 kpc along the SGP). The cosmic ionizing intensity is taken from Weymann et al (2001) but this is of the same order as the hot corona ($\varphi_\star\lesssim 10^{3.5}$ phot cm$^{-2}$ s$^{-1}$ at 75 kpc).
An earlier model attempted to explain the emission in terms of the
Stream's interaction with the halo (Bland-Hawthorn et al 2007). The 
direct interaction of the clouds
with the coronal gas is too weak to generate sufficient emission through
collisional processes, but these authors show that a `shock cascade'
arises if sufficient gas is stripped from the clouds such that the
following clouds collide with the ablated material.
This can be made to work if the Stream is moving through comparatively
dense coronal material ($n_{\rm hot}\sim 10^{-4}$ cm$^{-3}$).  But the
greater Stream distance ($D\gta75$ kpc; e.g. Jin \& Lynden-Bell 2008) makes
this less likely (Tepper-Garcia et al 2015). Barger et al (2017) adopt a massive hot halo in order to maximise the contribution from the shock cascade; whether such a corona is possible is still an open question (cf. Faerman et al 2017; Bregman et al 2018). 

The shock cascade model as presented above struggles to produce a Stream H$\alpha$ background of $\sim 100$ mR although there are other factors to consider in future models. The respective roles of magnetic fields (Konz et al 2001), thermal conduction (Vieser \& Hensler 2007) and turbulent mixing (Li et al 2019) have not been considered together in a dynamic turbulent boundary layer. They can work for or against each other in amplifying the observed recombination emission. Radiative/particle MHD models on galactic scales are in their infancy (Sutherland 2010; Bland-Hawthorn et al 2015) but will need to be addressed in future years.

\begin{figure*}[hbtp]
\centering
\includegraphics[scale=0.3]{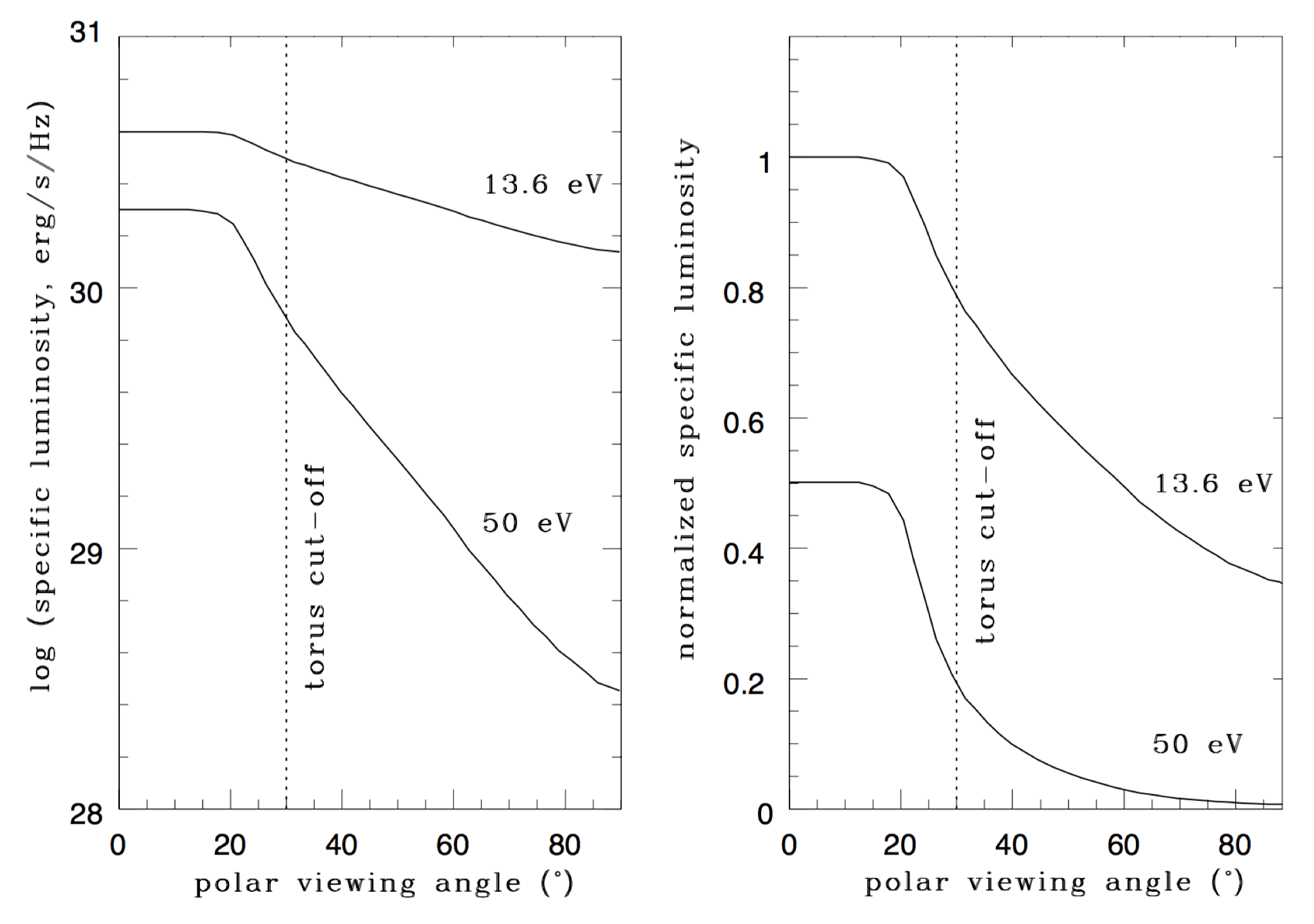}
\caption{Accretion disk model (Madau 1988; Madau et al 2014) for high mass accretion rates. The thick disk is defined within $r \lta 500\: r_g \approx 20$ AU, where $r_g = GM_\bullet/c^2$ is the black hole gravitational radius. It
produces an ionizing radiation field that is strongly
dependent on viewing angle and photon energy. The vertical dashed line indicates the cut-off imposed by the dusty torus
on much larger physical scales. 
(Left) Specific luminosity as a function of angle from
the SGP evaluated at two different photon energies.
(Right) The same model as the LHS now plotted on a linear scale,
normalized at the Lyman limit, to emphasize the 
self truncation of the disk radiation field, particularly at higher energies.
}
\label{f:Madau}
\end{figure*}

\begin{figure*}[hbtp]
\centering
\includegraphics[scale=0.4]{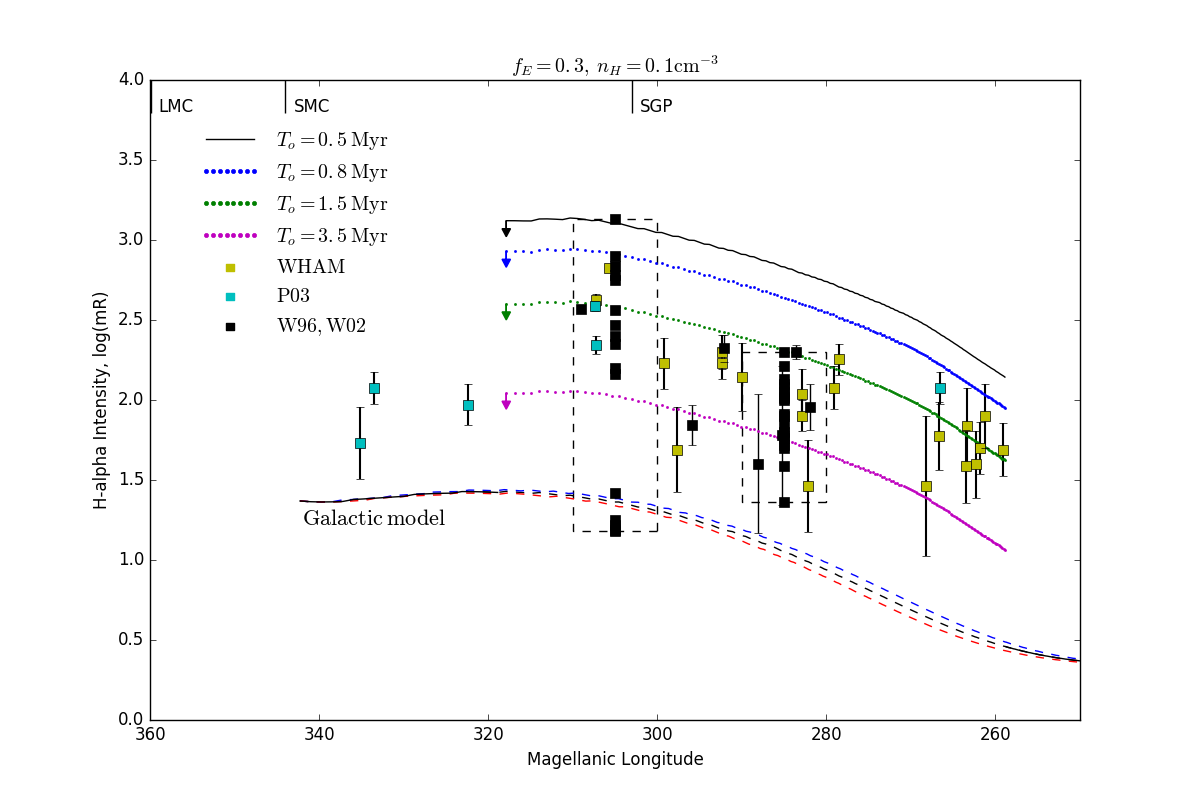}
\includegraphics[scale=0.4]{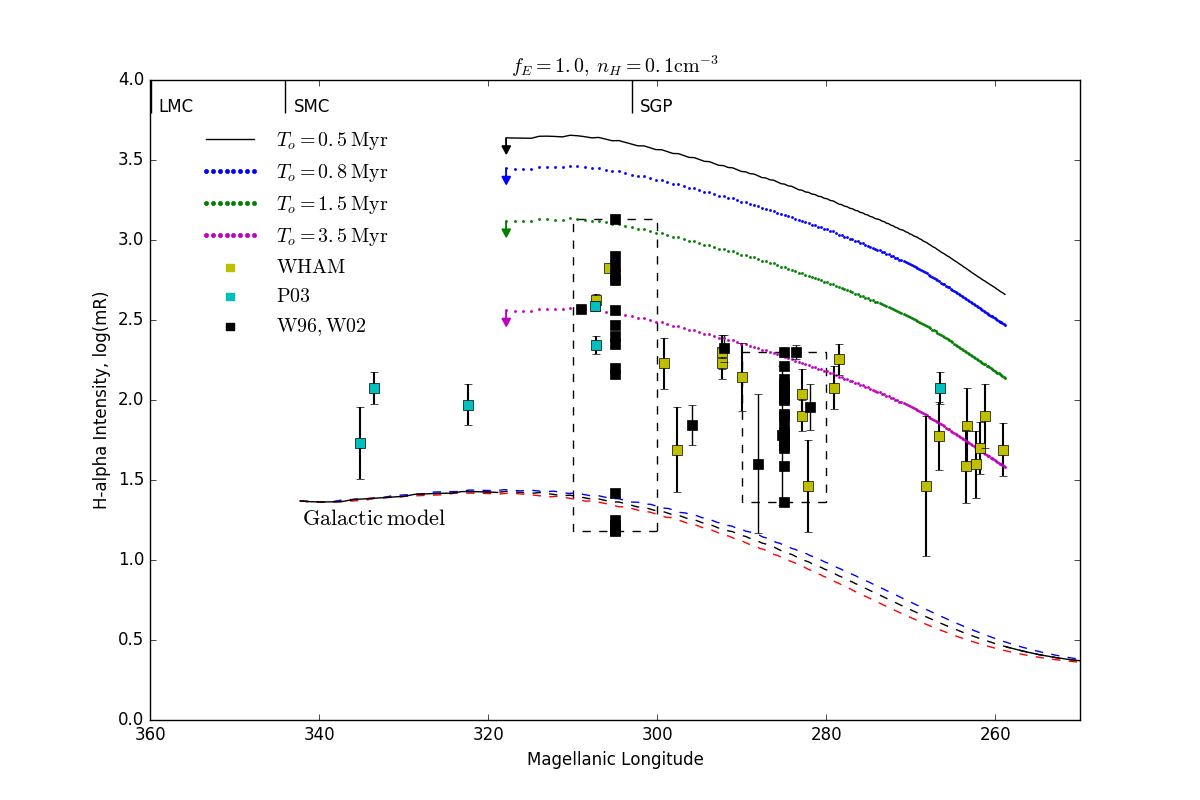}
\caption{The predicted H$\alpha$ intensity along
the Magellanic Stream ($D=75$ kpc)
as a function of Magellanic longitude.
The data points are taken from Weiner \& Williams (1996; W96),
Weiner et al (2002; W02), WHAM Survey (BH2013), and Putman et al (2003; P03). W96 and W02 are small
aperture measurements within a 10\deg\ window at unpublished sky positions.
The longitudes of the LMC, SMC and SGP are indicated. 
The topmost continuous black track corresponds to the middle track 
shown in Fig.~\ref{f:Seyfert} for $f_E=0.3$; this is the
instantaneous H$\alpha$ emission in the flash at $T_o = 0$ for optically thick
gas.
But since Sgr A* is in a dormant state, what we see today must have faded from 
the initial flash. We show the predicted trend for the
H$\alpha$ emission after 0.8 and 1.5 Myr (which includes 0.5 Myr for the light crossing time
to the Stream and back)
for an assumed density of $n_H$ $=$ 0.1 cm$^{-3}$. The
density cannot be much lower if we are to produce the 
desired H$\alpha$ emission; a higher density reduces the 
fading time.
The downward arrows indicate where the in-cone predicted emission
drops to the Galactic model outside of the cones. This is
shown as dashed lines along the bottom. 
}
\label{f:WHAM}
\end{figure*}

\subsection{Seyfert ionization model}

If the Galaxy went through a Seyfert phase in the recent past, it could conceivably have been so UV-bright that it lit
up the Magellanic Stream over the SGP through
photoionization (Fig. 4). The Magellanic Stream has detectable H$\alpha$ emission along its length five times more luminous than can be explained by UV escaping from the Galactic stellar population or an earlier starburst (BH2013, Appendix B).
The required star-formation rate is at least two orders of magnitude larger than allowed by the recent star formation history of the Galactic Centre (see \S 2). An accretion flare from Sgr A$^*$ is a much more probable candidate for the ionization source because (a) an accretion disk converts gas to ionizing radiation with much greater efficiency than star formation, thus minimizing the fuelling requirements; (b) there is an abundance of material in the vicinity of Sgr A* to fuel such an outburst.

We now consider the impact of past Seyfert activity using arguments that are independent of the x-ray/$\gamma$-ray mechanical timescales (\S 1), but consistent with them. We derive the likely radiation field of an accretion disk around a supermassive
black hole. BH2013 show how a Seyfert flare with an AGN spectrum that is 10\% of 
the Eddington luminosity ($f_E=0.1$) for a $4\times 10^6$M$_\odot$ black hole can produce sufficient UV
radiation to ionize the Magellanic Stream ($D \gtrsim 50$ kpc). But since Sgr A* is quiescent today, what we see has faded significantly 
from the original flash. H$\alpha$ recombines faster than the
gas cools for realistic gas densities ($n_e \sim 0.1-1$ cm$^{-3}$) and the well established Stream metallicity
($Z\approx 0.1\: Z_\odot$; Fox et al 2013). Thus, 
they find the
event must have happened within the last few million years, consistent 
with jet-driven models of the 10 kpc bipolar bubbles. 
This timescale
includes the double-crossing time (the time for the flare radiation
to hit the Stream $+$ the time for the recombination flux
to arrive at Earth), the time for the ionization front to 
move into the neutral gas and the recombination time.

\smallskip\noindent
{\sl Accretion disk model.}
The Shakura-Sunyaev treatment 
for sub-critical accretion produces a thin Keplerian disk that can cool on an infall timescale, leading to
a wide-angle thermal broadband emitter. 
They assumed an unknown source of turbulent stress 
generated the viscosity, e.g. through strong shearing
in the disk. 
But magnetorotational instability has supplanted hydrodynamical turbulence because even a weak magnetic field threaded through the disk suffices to trigger the onset
of viscosity (Balbus \& Hawley 1991).
The maximum temperature of the thin disk is
\begin{equation}
T_{\rm max}(r) \approx 54\:(r/r_s)^{-3/4} M_{\bullet,8}^{-1/4} f_E^{1/4} \ \ {\rm eV}
\end{equation}
where $M_{\bullet,8}$ is the mass of the black hole in units of 10$^8$\Msun\ (Novikov \& Thorne 1973). Thus the continuum
radiation peaks above 100 eV for a sub-critical accretion disk orbiting the black hole in Sgr A* which is sufficiently hardened
to account for the anomalous Stream ionization observed
in UV absorption lines (Fox et al 2014). Strictly speaking, 
$T_{\rm max}$ is for a maximally rotating (Kerr) black hole;
we need to 
halve this value for a stationary black hole.

In order to account for the mechanical luminosity of the x-ray/$\gamma$-ray bubbles, various authors (e.g. Zubovas \& Nayaksin 2012; Nakashima et al 2013) argue for an even more powerful
outburst of order the Eddington luminosity ($f_E \approx 1$). A quasi- or super-Eddington event in fact helps
all aspects of our work. This is more likely to 
generate sufficient UV to explain the Stream's 
ionization while providing sufficient mechanical
luminosity to drive a powerful jet or wind.
But the Shakura-Sunyaev algebraic formalism breaks 
down at high mass accretion rates ($f_E > 0.3$)
forming a geometrically thick, radiation-supported torus. These develop a central
funnel around the rotation axis from which most of the radiation arises. In Fig 4, ihe radiation field and spectral
hardness now have a strong dependence on polar angle (e.g. Paczynski \& Wiita 1980; Madau 1988).
The hot funnels may help to accelerate material along collimated jets (Abramowicz \& Piran
1980) which could further harden the radiation field and constrict the ionization pattern.

In Fig.~\ref{f:Madau}, we adopt the thick accretion disk model of Madau (1988) that ventures into the domain of mildly
super-Eddington accretion rates. (A supplementary discussion of this model is provided by
Acosta-Pulido et al 1990.)
The specific intensity of the thick disk (in units
of erg cm$^{-2}$ s$^{-1}$ Hz$^{-1}$) is 
given by
\begin{equation}
4\pi I_\nu = 1.0\times 10^{-14} T_s^{11/4}[\beta/(1-\beta)]^{1/2} x^{3/2} e^{-1}(1-e^{-x})^{-1/2}
\end{equation}
where $x=h\nu/kT_s$, $\beta$ is the ratio 
of the gas pressure to the total pressure ($\sim 10^{-4}$), and $T_s$ is the disk surface
temperature which has a weak dependence on the 
black hole mass and other factors, i.e.
$T_s \propto M_\bullet^{-4/15}\beta^{-2/15}$.
This parametric model
allows us to compute the ionizing spectrum for different viewing orientations of the disk. 

The most important attribute of an accretion disk model for our work is the photon flux and primary geometric parameters (e.g. inner and outer cut-off radii), with other considerations like spectral shape being of secondary importance (Tarter 1969; Dove \& Shull 1994; Maloney 1999). Madau (1988) includes a correction for scattering off the inner funnel which tends to harden the ionizing spectrum and boost its intensity. But it does not necessarily generate highly super-Eddington luminosities due to advection of heat onto the black hole (Madau et al 2014, Fig. 1). In \S 4.3, we consider a broad range of ionizing continua to uncover how the spectral hardness in the 10$-$100 eV window influences the predicted UV diagnostics.

\section{UV ionization of the Magellanic Stream}

\subsection{Expected emission from an active nucleus}
An accreting black hole
converts rest-mass energy with an efficiency factor
$\epsilon$ ($\sim 10\%$) into radiation with a luminosity $L_\bullet = \epsilon \dot m c^2 = 2\epsilon G \dot{m} M_\bullet/ r_s$,
for which $\dot m$ is the mass accretion rate and $r_s$ is
the Schwarzschild radius; for a recent review, see Zhang \& Lu (2019).  The accretion disk
luminosity can limit the accretion rate through radiation
pressure; the Eddington limit is given by
$L_E = 4\pi G M_\bullet m_p c\sigma_T^{-1}$
where $\sigma_T$ is the Thomson
cross-section for electron scattering. For the condition
$L_\bullet = L_E$,
radiation pressure from the accretion disk at the Galactic Centre
limits the maximum accretion rate to $\dot m \sim 0.2$ M$_\odot$
yr$^{-1}$.  Active
galactic nuclei appear to spend most of their lives operating at a
fraction $f_E$ of the Eddington limit with rare bursts arising from
accretion events (Hopkins \& Hernquist 2006). The orbital period of the Magellanic System is of order a Hubble time (Besla et al 2012) 
so we can consider the Stream to be a
stationary target relative to ionization timescales.

BH2013 show that for an absorbing cloud that is optically thick, the ionizing flux can be related directly to an H$\alpha$ surface brightness. The former is given by
\bee
\varphi_\bullet = 1.1\times 10^6 \left({{f_E}\over{0.1}}\right)
\left({f_{\bullet,{\rm esc}}}\over{1.0} \right)\left(D\over{\rm 75\
  kpc}\right)^{-2} \ \ \ {\rm phot\ cm}^{-2}\ {\rm s}^{-1} . 
\label{e:phi_agn}
\eee
The dust levels are very low in the Stream, consistent with its low metallicity (Fox et al 2013). We have included a term for the UV escape fraction from the AGN
accretion disk $f_{\bullet,{\rm esc}}$ (${\bf n}.{\bf N}=1$). 
The spectacular evacuated cavities observed at 21cm by
Lockman \& McClure-Griffiths (2016) suggest there is little
to impede the radiation along the poles, at least on large scales (Fig.~\ref{f:fermi}).
Some energy is lost due to Thomson scattering, but this is only a few percent in the best constrained sources (e.g. NGC 1068; Krolik \& Begelman 1988).  In principle, the high value of
$f_{\bullet,{\rm esc}}$ can increase $f_{\star,{\rm esc}}$ but the
stellar bulge is not expected to make more than a 10-20\% contribution
to the total stellar budget (Bland-Hawthorn \& Maloney 2002); a
possible contribution is accommodated by the factor $\zeta$
(equation~[\ref{f:Gal}]).

\begin{figure*}[htbp]
   \centering
  \includegraphics[scale=0.35]{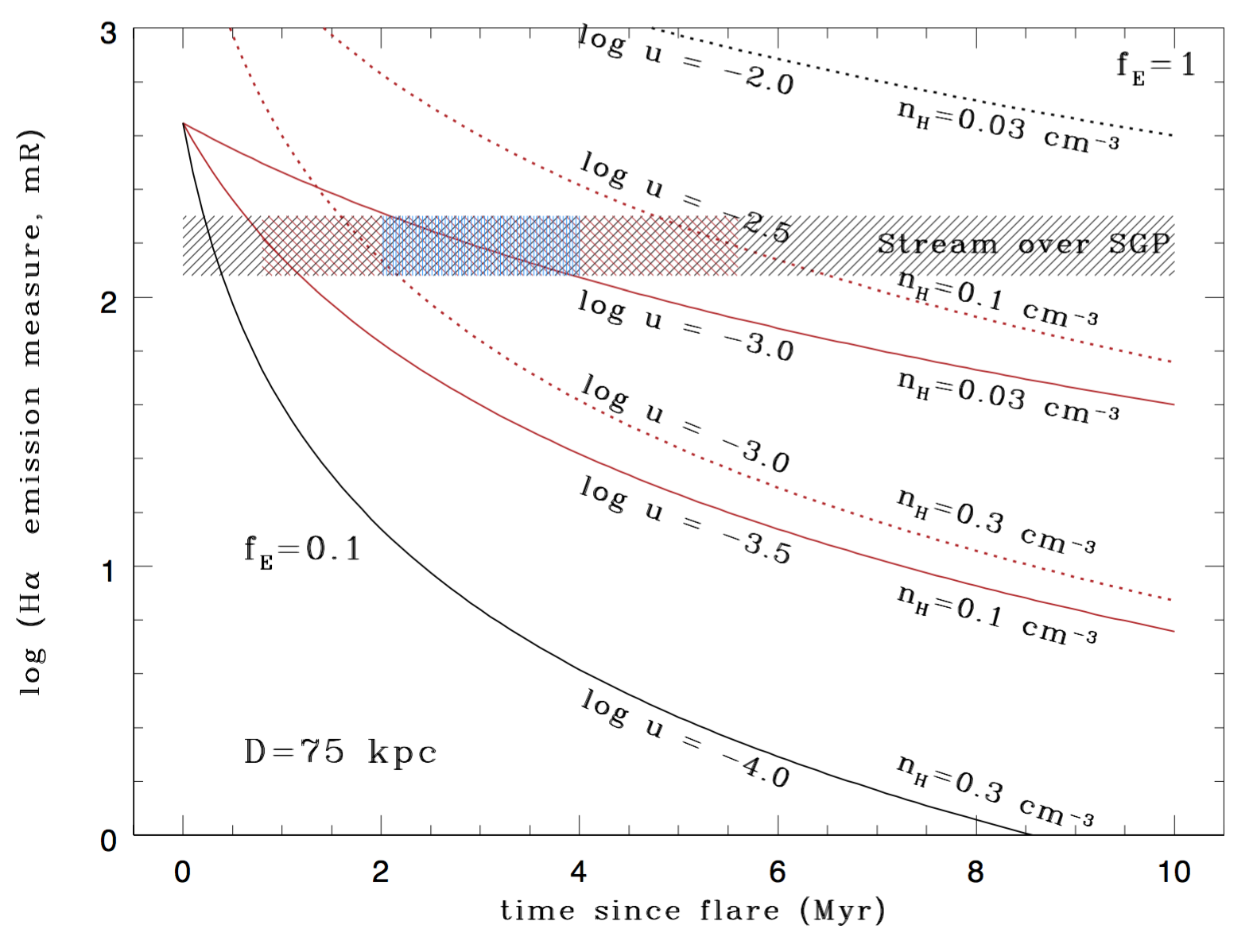}
  \caption{The decline in the \Ha\ surface brightness with
  time since the Seyfert flare event for gas clouds at
  a distance of $D=75$ kpc. The event ends abruptly in time
  at zero and the recombination signal declines depending on
  the cloud density. The light travel time there and back (roughly 0.5 Myr; BH2013) is not included here.
 Three tracks are shown (solid lines) for an Eddington fraction $f_E=0.1$ representing gas ionized at three different densities ($0.03-0.3$ cm$^{-3}$).
  The dotted lines show three tracks for an Eddington
  fraction of $f_E=1$. The value of the ionization parameter is shown at the time
  of the flare event. The hatched horizontal band is
  the observed \Ha\ surface brightness over the SGP.
  The red tracks are plausible models that explain the \Ha\ emission and these all fall within the red hatched region. The denser hatching in blue is a shorter duration fully consistent with timescales derived from the UV diagnostics (see text). Consistency between the independent diagnostics argues for $\log u=-3$ (independent of $f_E$) as characteristic of the \Ha\ emission along the Stream.
  }
   \label{f:fading}
\end{figure*}

The expected surface brightness for clouds that lie within an
`ionization cone' emanating from the Galactic Centre is given by
\bee
\mu_{\bullet,{\rm H}\alpha} = 440 \; \left({{f_E}\over{0.1}}\right)
\left({f_{\bullet,{\rm esc}}}\over{1.0} \right)\left(D\over{\rm 75\
  kpc}\right)^{-2} \ \ \ {\rm mR} . 
\label{e:mu_agn}
\eee
This provides us with an upper limit or `peak brightness' along the spin axis of the accretion disk assuming our model is correct. A few of the clumps exceed $\mu$(H$\alpha$) $\approx$ 440 mR in Fig. 6 by about a factor of 2. Our model parameters are only approximate.

Equation~\ref{e:mu_agn} is also applicable to isotropic emission
within the ionization cone from an unresolved point source
if the restriction is caused by an external screen,
e.g. a dusty torus on scales much larger than the accretion disk.
But here we consider thick accretion disk models that have 
highly angle-dependent radiation fields. This is evident for
Madau's radiation model in Fig.~\ref{f:Madau} with its footprint 
on the 
halo ionizing field shown in \ref{f:Seyfert}. Here the obscuring
torus has a half-opening angle $\theta_T = 30^\circ$; the 
accretion disk isophotes are 
seen to taper at $\theta_A = 20^\circ$. Both of
these values are illustrative and not well constrained by present the observations.

\subsection{Time-dependent analytical model of H recombination}

Thus far, we have assumed that some finite depth on the outer surface of a distant gas cloud comes into ionization equilibrium with the impinging radiation field. But what if the source of the ionizing radiation fades with time, consistent with the low Eddington fraction inferred today in the vicinity of Sgr A*? Then the ionization rate will decrease from the initial value for which equilibrium was established. We can treat the time-dependence of the H recombination lines analytically (BH2013); due to the presence of metal-line cooling, the C and Si ions require a more complex treatment with the time-dependent {\sl Mappings V} code. This analysis is covered in the next section where we find, in fact, that the \Ha\ and UV diagnostics arise in different regions of the Magellanic Stream.

After Sharp \& Bland-Hawthorn (2010), we assume an exponential decline for $\varphi_i$, with a characteristic timescale for the ionizing source $\tau_s$. The time-dependent equation for the electron fraction $x_e = n_e/n_H$ is
\bea
{dx_e\over dt} &=& -\alpha n_H x_e^2 + \zeta_0 e^{-t/\tau_s}(1-x_e)
\label{e:dxdt}
\eea
where $\zeta$ is the ionization rate per atom. This was solved for in BH2013 (Appendix A). If we let $\tau_s \rightarrow 0$, so that $\varphi_i$
declines instantaneously to zero, we are left with
\bee
{dx_e\over dt} = -\alpha n_H x_e^2
\eee
For the initial condition $x_e=1$ at $t=0$, we get
\bee
x_e = \left(1+t/\tau_{\rm rec}\right)^{-1}
\label{e:ionfrac}
\eee
for which the recombination time $\tau_{\rm rec} = 1/\alpha n_H$ and $\alpha_B = 2.6\times 10^{-13}$ cm$^3$ s$^{-1}$ for the recombination coefficient (appropriate for hydrogen at $10^4$ K). Thus the emission measure 
\bee
{\cal E}_m = 1.25 \varphi_6 x_e^2(t)\; {\rm cm^{-6}\;pc}
\eee
where $\varphi_6$ is the ionizing photon flux in units of $10^6$ phot cm$^{-2}$ s$^{-1}$. It follows from equation~\ref{e:mu_agn} that
\bee
\mu_{\bullet}(t) = 440 \; \left({{f_E}\over{0.1}}\right)
\left({f_{\bullet,{\rm esc}}}\over{1.0} \right)\left(D\over{\rm 75\
  kpc}\right)^{-2} (1+t/\tau_{\rm rec})^{-2} \ \ \ {\rm mR} . 
\label{e:mu_agn_t}
\eee

\begin{figure}[htbp]
   \centering 
  \includegraphics[scale=0.43]{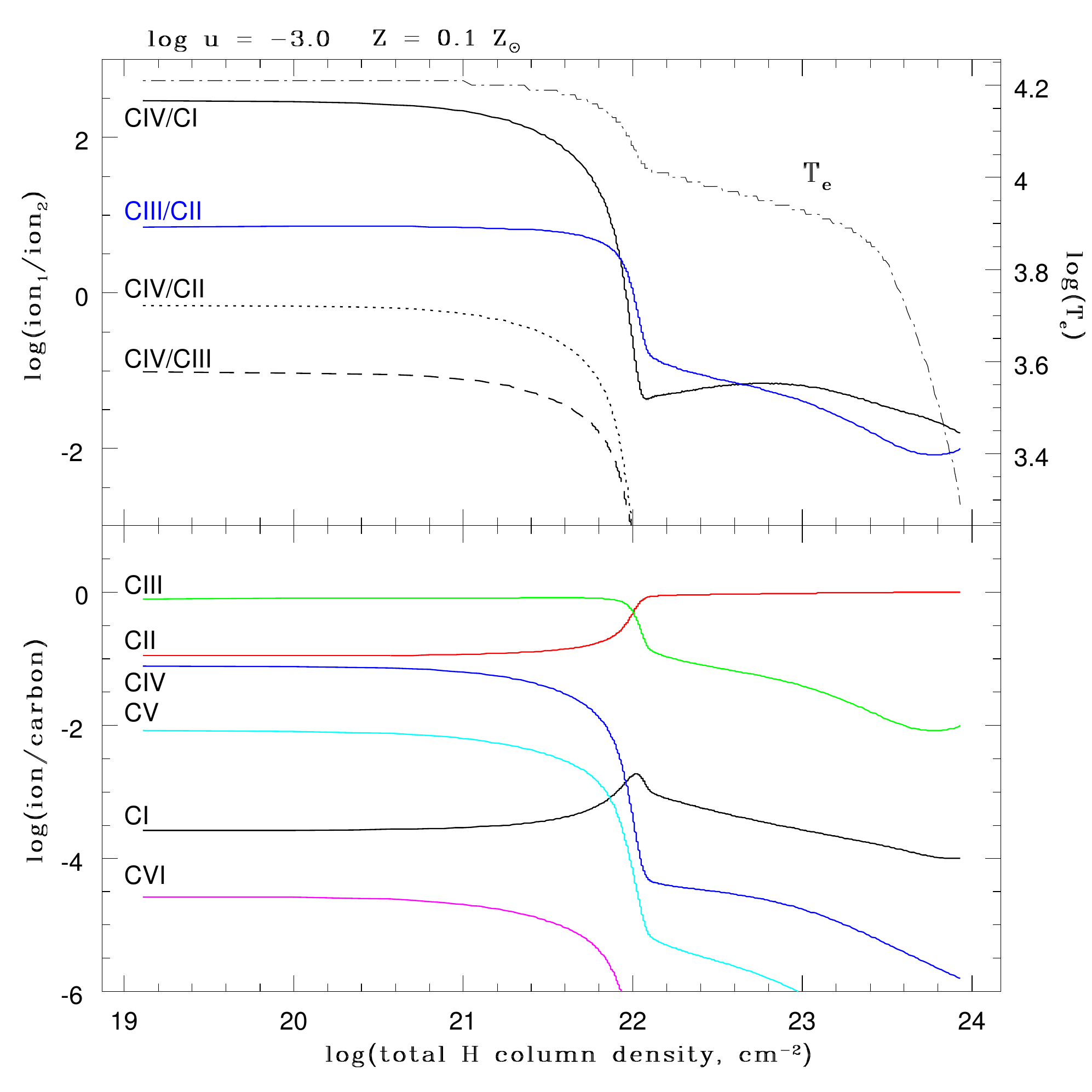}
  \caption{
  {\sl Mappings V} ionization calculation for a {\it continuously radiating} power-law source in Fig.~\ref{f:spec}(e) ($\alpha=-1$).  At the front face, the radiation hits a cold slab of gas with sub-solar metallicity ($Z=0.1 Z_\odot$) and ionization parameter, $\log u = -3.0$. In the upper panel, the change in the log ratio of two carbon (C) ions is shown as a function of depth into the slab. The log column densities on the horizontal axis are total H densities (\HI+\HII) and are much larger than for the fading models. The dot-dashed line is the electron temperature $T_e$ of the gas as a function of depth, as indicated by the RHS. The lower panel shows the log ratio of each C ion to the total carbon content as a function of depth. The \CIV/\CII\ model track is to be compared to the data points in Fig.~\ref{f:CIV}.
 }
   \label{f:carbon}
   \medskip
\end{figure}


\begin{figure*}[htbp]
   \centering
  \includegraphics[scale=0.4]{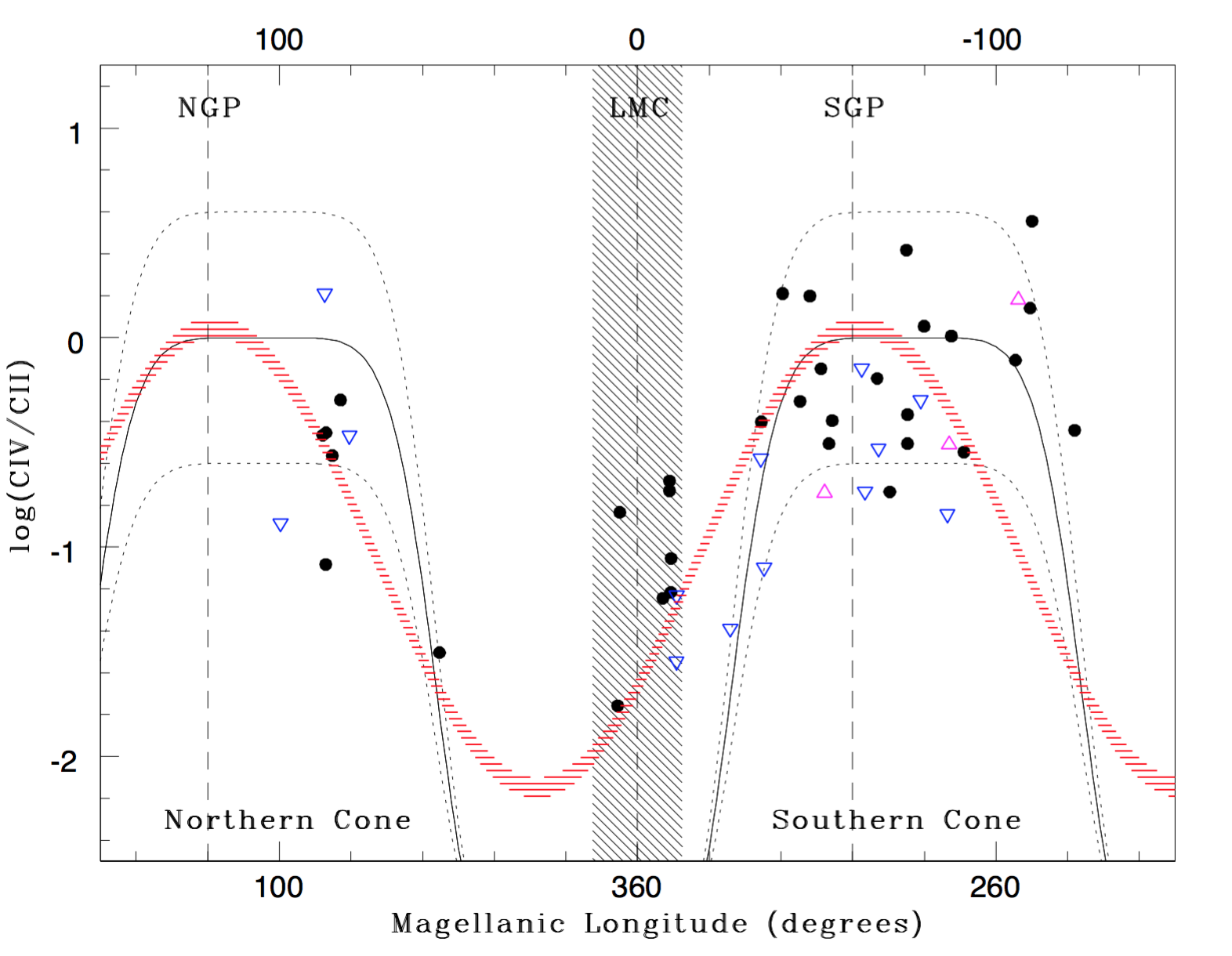}
\caption{The column-density ratio of \CIV/\CII\ (Fox et al 2014) along the Magellanic Leading Arm (left of shaded band) and trailing Stream (right of shaded band) presented as a function of Magellanic longitude $\ell_{\rm M}$. Detections are shown as solid symbols with typical $1\sigma$ errors being twice the size of the symbol; upper limits are shown as blue triangles, lower limits as magenta triangles. The NGP, LMC longitude and SGP are all indicated as vertical long-dashed lines. The measured values within the shaded vertical band fall along the LMC sight line. 
The domain of the ionization cones (NGP and SGP; see Fig. 2) is indicated by the two $\cap$-shaped curves; the dotted lines indicate $\pm 0.25$ dex in $\log u$. The specific Madau accretion disk model used is discussed in \S 3.
Note that some of the enhanced \CIV/\CII\ values -- seen against the
`leading arm' of the Stream -- fall within the NGP cone. The slightly elevated values in the LMC's vicinity may be due to hard (e.g. ULX) ionizing sources within the dwarf galaxy.
The red sinusoid (\S 5) is an attempt to force-fit the distribution of \CIV/\CII\ line ratios with spherical harmonics as a function of the sky coordinates. 
}
   \label{f:CIV}
   \medskip
\end{figure*}

Equations (\ref{e:mu_agn}) and (\ref{e:mu_agn_t}) have several important implications. Note that the peak brightness of the emission depends only on the AGN parameters and the Stream distance, not the local conditions within the Stream. (This assumes that the gas column density is large enough to absorb all of the incident ionizing flux, a point we return to in \S \ref{s:critical}.) Hence, in our flare model, the Stream gas just before the ionizing photon flux switches off may not be uniform in density or column density, but it would appear uniformly bright in \Ha. After the ionizing source turns off, this ceases to be true: the highest-density regions fade first, because they have the shortest recombination times; the differential fading scales as $1/(1+t/\tau_{\rm rec})^2$. This is clearly seen in BH2013 (Fig. 6) which shows the \Ha\ surface brightness versus $n_e$ for fixed times after the flare has ended: at any given time, it is the lowest density gas that has the brightest \Ha\ emission, even as all of the Stream is decreasing in brightness.

In Fig.~\ref{f:fading}, we show two sets of three fading curves defined by two values of $f_E$, 0.1 and 1, the range suggested by most AGN models of the x-ray/$\gamma$ ray bubbles (e.g. Guo \& Mathews 2012), although higher super-Eddington values have been proposed (e.g. Zubovas \& Nayakshin 2012). The three curves cover the most likely range of cloud density $n_H$ (derived in the next section). The model sets overlap for different combinations of $f_E$ and $n_H$. The hatched horizontal band is the median \Ha\ surface brightness over the SGP as discussed in BH2013. The horizontal axis is the elapsed time since the Seyfert flare event.

The red tracks are reasonable models that explain the \Ha\ emission and these all fall within the red hatched region. 
The denser hatching in blue is a more restricted duration to explain the \CIV/\CII\ values at the SGP (\S 4.4). With the UV diagnostic constraints from {\it HST}, we find that $\log u_o \sim -3$ is a reasonable estimate of the initial ionization parameter that gave rise to the \Ha\ emission we see today. In Fig.~\ref{f:carbon}, we see that a {\it continuous} radiation field fixed at $u_o$ can produce the UV diagnostics observed but such models require very large \HI\ column densities (\S 4.3). This situation is unrealistic given the weak AGN activity observed today.  Thus, to accommodate the fading intensity of the source, we must start at a much higher $u$ to account for the UV diagnostics. Below, we find that the observed C and Si absorption lines are unlikely to arise from the same gas that produces \Ha.

\subsection{Critical column density associated with flare ionization}
\label{s:critical}
We have assumed until now that the Magellanic Stream has sufficient hydrogen everywhere within the observed solid angle to absorb the ionizing UV radiation from the Seyfert nucleus (e.g. Nidever et al 2008). For a continuous source of radiation (e.g. Fig.~\ref{f:carbon}), this requires an H column density greater than a critical column density \Ncr\ given by
\be
\Ncr \approx 3.9\times 10^{19} \phi_6 (\nH/0.1)^{-1}\;\; {\rm cm}^{-2}
\label{e:Ncrit0}
\ee
where $\phi_6$ is the ionizing UV luminosity in units of 10$^6$ phot cm$^{-2}$ s$^{-1}$. For simplicity, we set $D=75$ kpc and $f_{\rm esc}=1$. By substituting from equation~\ref{e:phi_agn}, we find
\be
\Ncr \approx 4.2\times 10^{20}\; (\fE/0.1)(\nH/0.1)^{-1}\;\; \rm{cm^{-2}} 
\label{e:Ncrit1} 
\ee
where \fE\ is the Eddington fraction and \nH\ is the local hydrogen volume density in units of cm$^{-3}$. Thus
\be
\Ncr \approx 1\times 10^{20}\; u_{-3}\;\; \rm{cm^{-2}}
\label{e:Ncrit2}
\ee
where $u_{-3}$ is the ionization parameter in units of $10^{-3}$, consistent with Fig.~\ref{f:fading}, and where it follows
\be
u_{-3} = 0.37 (\fE/0.1)(\nH/0.1)^{-1} .
\label{e:Ncrit3}
\ee
Barger et al (2017, Fig. 12) plot the \HI\ column densities (averaged over the same 1 degree beam as their WHAM \Ha\ observations) versus the \Ha\ intensity. The measured values suggest that the total HI column may fall below that set by equation~\ref{e:Ncrit1} except in high-density regions ($\nH \geq 1$). In this case, the peak \Ha\ surface brightness will be reduced by a factor $\sim N_{\rm H}/N_{\rm cr}$ from the value predicted by equation \ref{e:mu_agn}. This will contribute to, and could even dominate (see \S \ref{s:patchy}) the spread in the observed $\mu(\Ha)$. For lines of sight with $N_{\rm H} > N_{\rm cr}$, there is a constant `ionized column' recombination rate, balancing the incident ionizing flux. At the time of the Seyfert flash, once ionization equilibrium is reached (note that equations~\ref{e:Ncrit0} to \ref{e:Ncrit3} only apply when the central source is switched on), regions of lower gas density will extend deeper along the line-of-sight (and hence to larger \Np) to compensate for the lower \ne.

\begin{figure*}[htbp]
   \centering
  \includegraphics[scale=0.7]{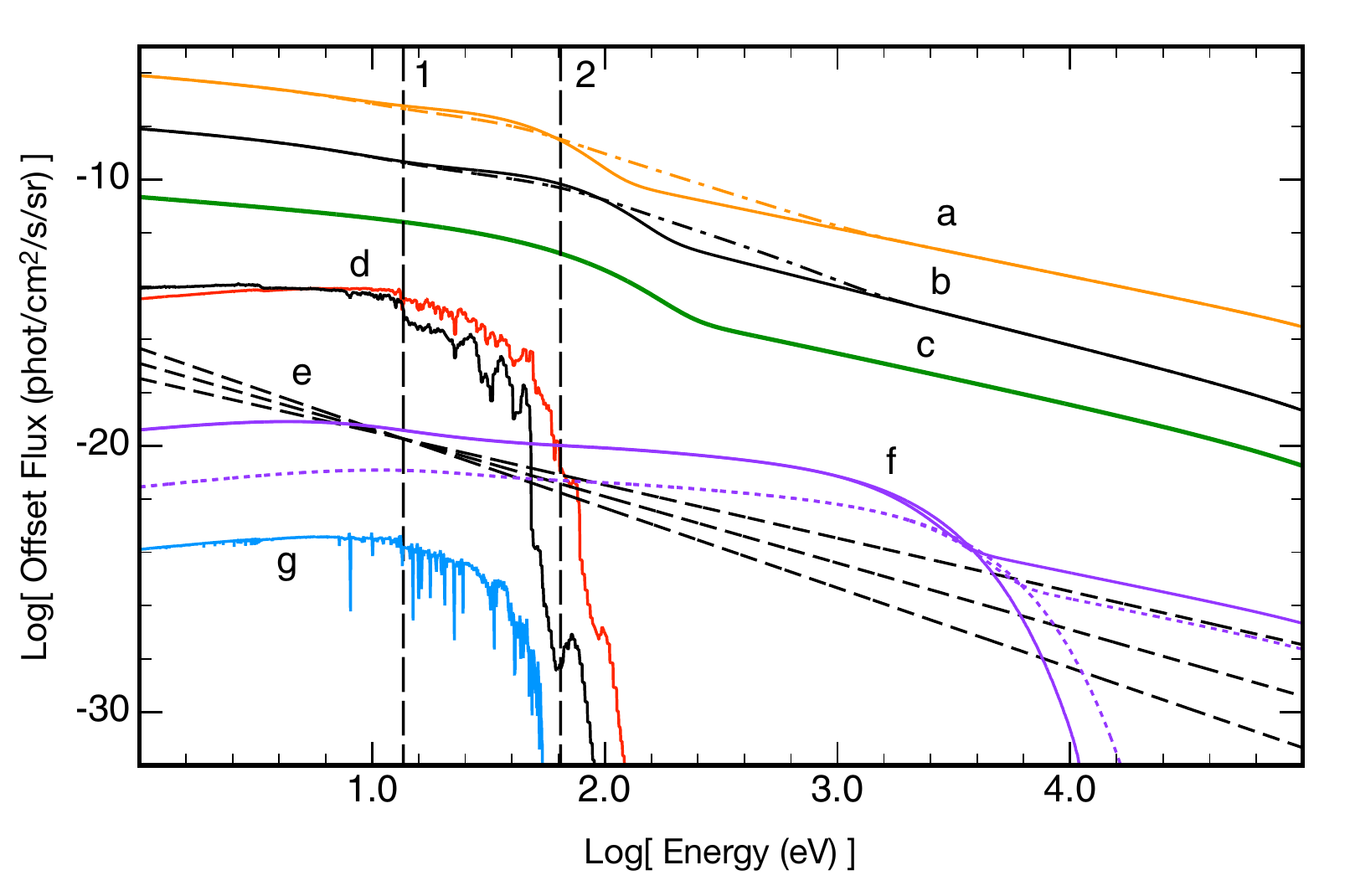}
  \caption{
 The broad distribution of ionizing continua explored within the current work using {\sl Mappings V}. The offsets along the vertical axis are arbitrary: all model spectra are normalized to the same photon number in the window indicated by the vertical dashed lines (1, 2), important for the production of H, Si and C ions. From top to bottom: generic (a) broad-line and (b) narrow-line Seyfert spectra from OPTXAGNF code (Done et al 2012; Jin et al 2012) where the dot-dashed line includes a 0.2 keV `soft Compton' corona $-$ both are scaled to $M_{\bullet}$ in Sgr A* ($R_c=60 R_g$, $f_{\rm PL}=0.4$, $\Gamma\approx 2$);  (c) Seyfert spectrum derived by JBH2013 from NGC 1068 observations; (d) Starburst99 spectra for impulsive burst (red) and extended 4~Myr phase (black) assuming a Kroupa IMF;
  (e) power-law spectra with $f_\nu \propto \nu^{\alpha}$ for which $\alpha=-1.0, -1.5, -2.0$;
  (f) a total of four ULX spectra from OPTXAGNF code split between $M_\bullet=100$\Msun\ (dotted line) and $M_\bullet=1000$\Msun\ (solid line), both models with an inner disk ($R_c=6 R_g$), and one case each with an extended component ($R_c=20 R_g$, $f_{\rm PL}=0.2$, $\Gamma\approx 2$) $-$ all models are fed for 1 Myr at $f_E=1$;
  (g) hot star from CMFGEN code with solar metallicity, surface temperature 41,000 K and surface gravity $\log\;g=3.75$ (Hillier 2012).}
   \label{f:spec}
\end{figure*}

\subsection{Time-dependent Mappings model of C, Si recombination}

We use the {\sl Mappings V} code  (Sutherland \& Dopita 2017)  to study the ionization, recombination and cooling of the C and Si ions at the surface of Magellanic Stream clouds. To determine the expected column depths of the different ionization states, we explore a broad range of {\sl Mappings V} photoionization models extending across black-hole accretion disk, starburst and individual stellar sources. The full range of models is illustrated in Fig.~\ref{f:spec}. The vertical dashed lines at 13.6 eV and 64.5 eV\footnote{The \CIV\ ionization potential is 47.9 eV but \CV\ at 64.5 eV is important for reducing \CIV\ in the presence of a hard ionizing continuum.} delimit the most important energy range in the production of the H, Si and C ions in our study.

For both the AGN and starburst/stellar photoionization models, we assume: (i) a constant density ionization-bounded slab with $n_{\rm H}=0.01$ cm$^{-3}$, (ii) a gas-phase metallicity of $Z=0.1 Z_{\odot}$ (Fox et al 2013) made consistent for all elements with concordance abundances (Asplund 2005); (iii) at these low metallicities, we can ignore depletion onto dust grains; (iv) we assume that the gas column density is large enough everywhere to absorb all of the incident ionizing flux (\S 4.3). The results are only weakly dependent on $n_{\rm H}$ but have a strong dependence on the ionization parameter $u=\varphi/(c n_H)=10^6\varphi_6/(c n_H)$, which is how we choose to discuss the main results.

The set up for all photoionization models is given in Table~\ref{t:models} where the required ionized and neutral column densities are listed in columns 4 and 5. In column 3, we show the instantaneous electron temperature $T_e$ after the flash occurs; values indicated in italics are generally too low for sustained enhancements in all of the UV diagnostics (ion ratios, column densities, etc.). We explore each of the models below but, in summary, we find that for a fading source, only the accretion-disk driven radiation fields at high ionization parameter ($\log u \gta -2$) generate the high temperatures required to reproduce the observed UV diagnostics. We show illustrative plots for each diagnostic below, but the results are tabulated in Appendix B (Tables 2-3).

\begin{figure*}[htbp]
   \centering 
  \includegraphics[scale=0.4]{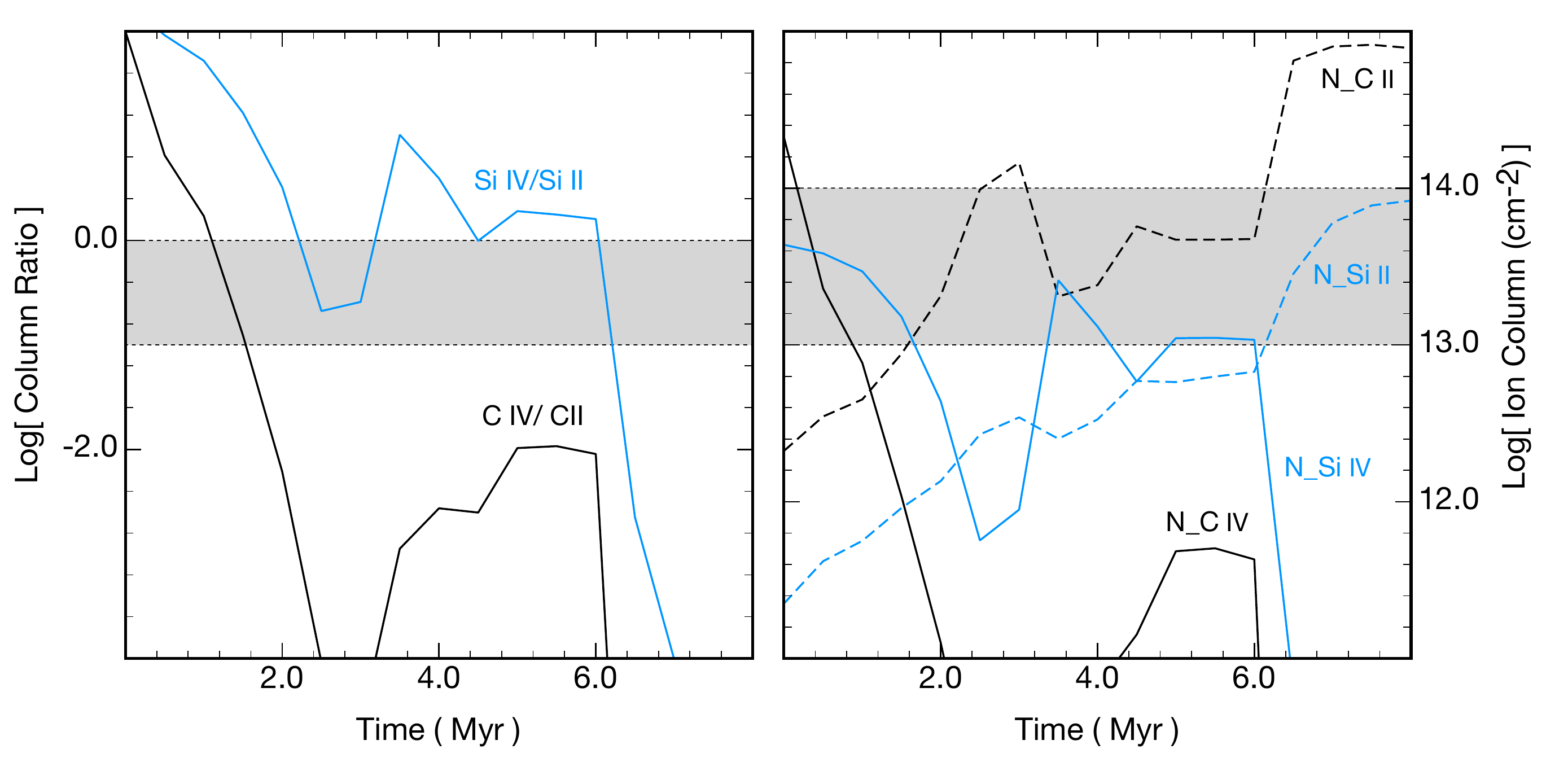}
  \includegraphics[scale=0.4]{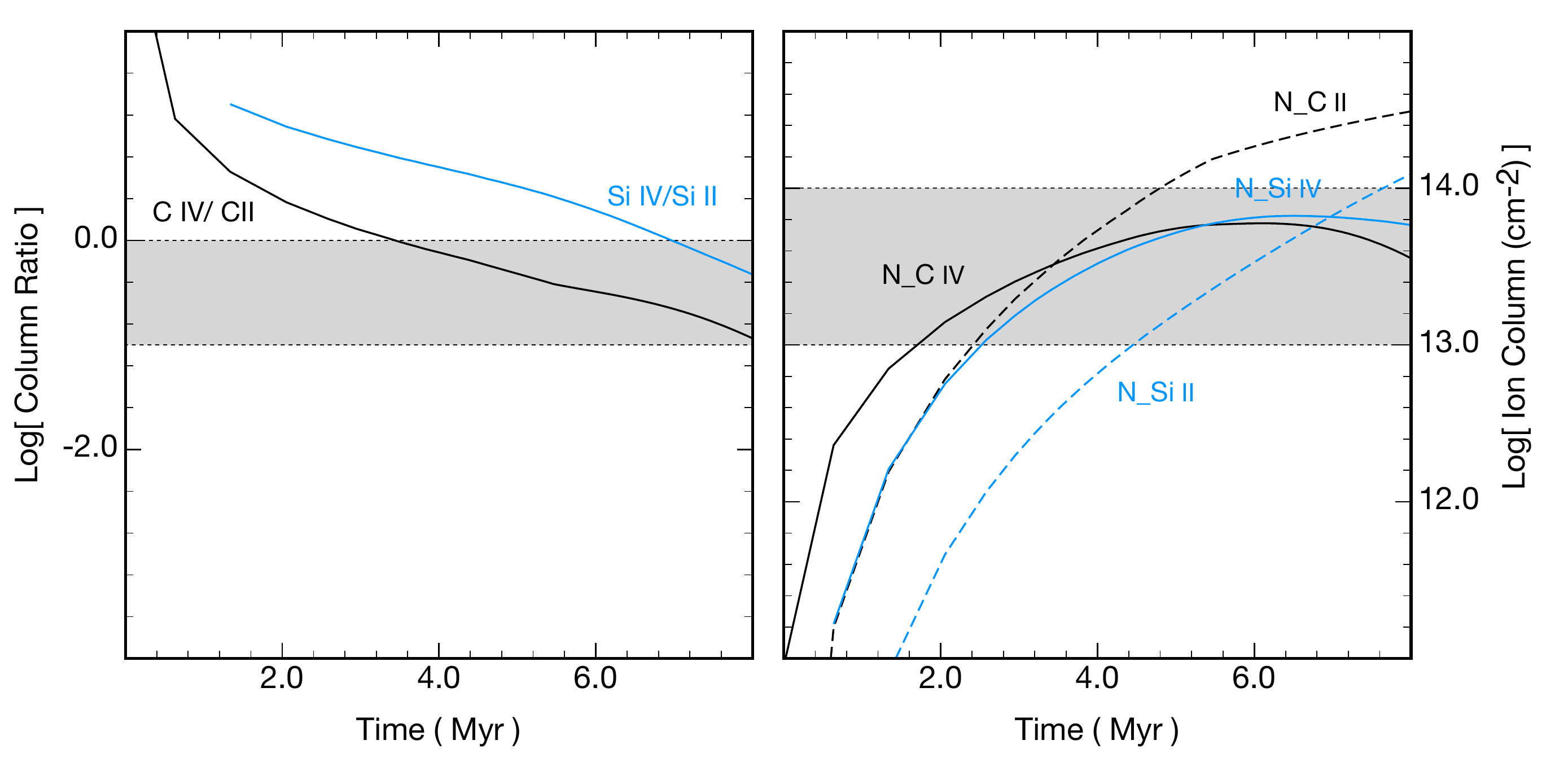}
  \caption{
  {\sl Mappings V} time-dependent ionization calculation (top) for the fading star cluster model in Fig.~\ref{f:spec}(d); (bottom) for the fading ULX model with the hardest spectrum in Fig.~\ref{f:spec}(f). We show the time evolution of the \SiIV/\SiII\ and \CIV/\CII\ ratios (left) and projected column density of all four ions (right).
  The light travel time there and back (roughly 0.5 Myr; BH2013) is not included here.
  The grey horizontal band encloses most of the high-ionization data points along the Magellanic Stream (Fig.~\ref{f:CIV}; Fox et al 2014). A stellar or starburst, photoionizing spectrum fails to produce sufficient \CIV\ or \SiIV\ absorption regardless of its bolometric luminosity. In principle, a ULX spectrum can produce the observed UV diagnostic ratios and column densities; this may account for the enhanced \CIV/\CII\ localised around the LMC (Fig.~\ref{f:CIV}). The initial ionization parameter at the front face of the slab is $\log u_o = -1$ for both models ($Z=0.1Z_\odot$). At $\log u_o=-2$, the tracks in the top figures fall below the grey band, and the tracks in the bottom figures cross the grey band in half the time. The results for more ions are presented in Appendix B.
 }
   \label{f:stellarfading}
   \medskip
\end{figure*}

\begin{figure*}[htbp]
   \centering 
  \includegraphics[scale=0.5]{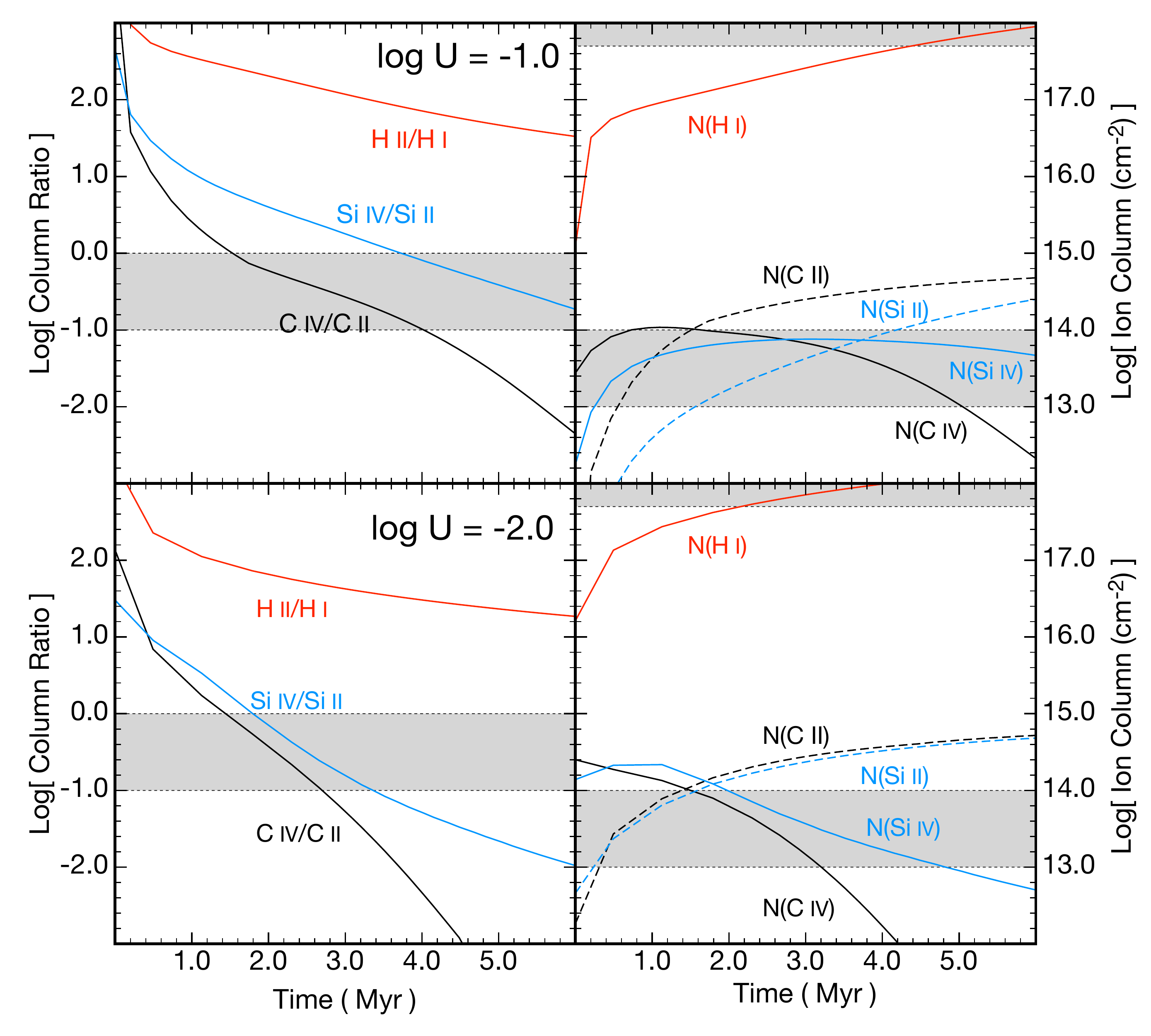}
  \caption{
 {\sl Mappings V} time-dependent ionization calculation for the fading AGN model in Fig.~\ref{f:spec}(c). The initial ionization parameter at the front face of the slab is $\log u_o = -1$ (upper) and $\log u_o = -2$ (lower) where $Z=0.1Z_\odot$.  On the LHS, the grey band refers to the observed \CIV/\CII\ and \SiIV/\SiII\ ratios; there is no UV constraint for \HII/\HI. The grey horizontal bands enclose most of the high-ionization data points along the Magellanic Stream (Fig.~\ref{f:CIV}; Fox et al 2014). The AGN models considered (Table~1) give essentially the same results with only small differences in the trends. On the RHS, the evolution in projected column density is shown for four metal ions and \HI\ determined from UV spectroscopy. The top grey band refers to \HI\ for which most values quoted in Fox et al (2014) are upper limits; the bottom grey bands refers to the metal ions. For the LHS tracks to fall within the grey band simultaneously, over the allowed range of $u_o$ ($-2 < \log u_o < -1$), the estimated time span is $2-4$ Myr. The light travel time there and back (roughly 0.5 Myr; BH2013) is not included here. The results for more ions are quantified in Appendix B.
 }
   \label{f:AGNfading}
   \medskip
\end{figure*}
\begin{table}[!ht]
\caption{Photoionization models: initial conditions in the gas slab when the ionizing source is switched on. We consider the internal ionization structure to be instantaneous ($\ll$ 1 Myr) because the ionization front propagates rapidly into the slab (BH2013). The columns are (1) ionizing source in Fig.~\ref{f:spec}, (2) ionization parameter, (3) initial electron temperature, (4) ionized H column, (5) neutral H column.  The temperatures in italics produce instantaneous highly ionized C and Si in the flash but are too low to produce sustained high ion ratios when the source fades. The predicted C and Si ion columns are given in Appendix B.}
\begin{center}
\begin{tabular}{ccccc}
\hline
Model & $\log\:u$ & initial $T_e$ & N(\HII) & N(\HI) \\ 
\hline
\hline
BLS (a) &-1.0&2.99E+04&3.08E+19&1.32E+15 \\ 
&-2.0&2.13E+04&3.08E+19&1.72E+16 \\ 
&-3.0&{\it 1.57E+04}&3.05E+19&2.59E+17 \\ 
BLS + soft Compton (a)  &-1.0&3.50E+04&3.08E+19&1.24E+15 \\ 
&-2.0&2.31E+04&3.08E+19&1.73E+16 \\ 
&-3.0&{\it 1.63E+04}&3.05E+19&2.82E+17 \\ 
\hline
NLS (b) &-1.0&3.26E+04&3.08E+19&1.59E+15 \\ 
&-2.0&2.26E+04&3.08E+19&2.12E+16 \\ 
&-3.0&{\it 1.61E+04}&3.05E+19&3.49E+17 \\ 
NLS + soft Compton (b) &-1.0&3.67E+04&3.08E+19&1.38E+15 \\ 
&-2.0&2.35E+04&3.08E+19&1.99E+16 \\ 
&-3.0&{\it 1.63E+04}&3.05E+19&3.39E+17 \\ 
\hline
BH2013 (c) &-1.0&3.18E+04&3.08E+19&1.26E+15 \\ 
&-2.0&2.24E+04&3.08E+19&1.65E+16 \\ 
&-3.0&{\it 1.61E+04}&3.05E+19&2.60E+17 \\ 
\hline
Star cluster (d) &-1.0&{\it 1.55E+04}&3.08E+19&1.14E+15 \\ 
&-2.0&{\it 1.50E+04}&3.08E+19&1.17E+16 \\ 
&-3.0&{\it 1.36E+04}&3.07E+19&1.49E+17 \\ 
\hline
PL, $\alpha=-1.0$ (e) &-1.0&4.11E+04&3.08E+19&9.69E+14 \\ 
&-2.0&2.43E+04&3.08E+19&1.50E+16 \\ 
&-3.0&{\it 1.65E+04}&3.06E+19&2.55E+17 \\ 
PL, $\alpha=-1.5$ (e) &-1.0&3.45E+04&3.08E+19&8.71E+14 \\ 
&-2.0&2.31E+04&3.08E+19&1.19E+16 \\ 
&-3.0&{\it 1.62E+04}&3.06E+19&1.87E+17 \\ 
PL, $\alpha=-2.0$ (e)  &-1.0&3.03E+04&3.08E+19&8.16E+14 \\ 
&-2.0&2.17E+04&3.08E+19&1.05E+16 \\ 
&-3.0&{\it 1.55E+04}&3.06E+19&1.59E+17 \\ 
\hline
ULX100 (f) &-1.0&3.70E+04&3.08E+19&7.49E+15 \\ 
&-2.0&2.07E+04&3.07E+19&1.47E+17 \\ 
&-3.0&{\it 1.43E+04}&2.50E+19&5.80E+18 \\ 
ULX1000 (f) &-1.0&4.00E+04&3.08E+19&4.55E+15 \\ 
&-2.0&2.19E+04&3.07E+19&8.83E+16 \\ 
&-3.0&{\it 1.50E+04}&2.73E+19&3.48E+18 \\ 
\hline
40,000 K Star (g) &-1.0&{\it 1.53E+04}&3.08E+19&1.13E+15 \\ 
&-2.0&{\it 1.49E+04}&3.08E+19&1.16E+16 \\ 
&-3.0&{\it 1.36E+04}&3.07E+19&1.48E+17 \\ 
\hline
\end{tabular}
\end{center}
\label{t:models}
\end{table}

\subsubsection{Stellar, starburst and ULX models}
In Fig.~\ref{f:stellarfading} (top), 
even though star forming regions in either the LMC or the Galaxy do not contribute significantly to the ionization of the Magellanic Stream, for completeness, we include a {\it Mappings V} time-dependent ionization calculation for the fading star cluster model in Fig.~\ref{f:spec}(d). We present the evolution of the \SiIV/\SiII\ and \CIV/\CII\ ratio (left) and the evolution in projected column density of \HI\ and all four metal ions (right). The grey horizontal band encloses most of the data points along the Magellanic Stream (Fox et al 2014). A comparison of both figures shows that the gas layer is cooling down through metal-line (and H) recombination. A stellar or starburst photoionizing spectrum fails to produce sufficient \CIV\ or \SiIV\ absorption (Tables 1-3), regardless of its bolometric luminosity.

For the incident ionizing radiation field, we also explore the CMFGEN O-star grid of Hillier (2012), and settle on an O-star with $T_{\rm eff}=41000$ K and $\log\:g=3.75$, which represents a somewhat harder version of the typical Milky Way O-star. Importantly, this ionizing spectrum is unable to excite appreciable amounts of \CIV\ or \SiIV. The same holds true for static photoionization models.
Lower $u$ values ($\log\,u < -2.0$) and stellar photoionization both
fall short of producing such high columns and column ratios that, taken together, are a serious challenge for any model. Typical \CIV\ columns from the hard stellar spectra rarely exceed $10^{10}$ cm$^{-2}$ for a reasonable range of $u$. 

But there is a special case we need to consider that is not factored
into the existing Starburst99 models. Ultraluminous x-ray sources (ULX) are known to be associated with vigorous star-forming regions and, indeed, a few have been observed in the LMC (Kaaret et al 2017). In Fig.~\ref{f:stellarfading}, we show that the hard spectrum of the ULX source can in principle achieve the 
observed UV diagnostics along the Magellanic Stream. We do not believe one or more ULX sources account for the enhanced values over the SGP although they could account
for the slightly elevated levels observed near the LMC (Fig.~\ref{f:CIV}).
There are numerous problems with an LMC explanation as has been explored 
in earlier work. Barger et al (2013) show that the mutual ionization of
the LMC and SMC on their local gas is well established, as are their respective
orientations. The \Ha\ surface brightness declines with radius for both sources. Furthermore, the \CIV/\CII\ and \SiIV/\SiII\ ratios rise
dramatically as we move {\it away} from the LMC in Magellanic longitude $\ell_M$ (Fig.~\ref{f:CIV}) which is entirely inconsistent for the LMC being responsible. The extent of the LMC ionization
is illustrated in Fig.~3.

\subsubsection{AGN models}
\label{s:agn}

The bursty stellar ionizing radiation from the LMC or from the Galaxy fails by two orders of magnitude to explain the Stream (BH2013, Appendix B). We believe the most reasonable explanation is the fading radiation field of a Seyfert flare event.
In Fig.~\ref{f:fading}, the incident AGN radiation field strength is defined in terms of the initial ionization parameter $u$ and explore 3 tracks that encompass the range expected across the Magellanic Stream: $\log\,u=-3.5, -3.0, -2.5$. 
As argued in \S 4.2, this range can account for the Stream \Ha\ emissivity but the UV signatures likely arise under different conditions.
We now investigate the C and Si diagnostics because of their potential to provide an independent estimate of when the Seyfert flare occurred.

Here, we explore a wide range of models summarised in Fig.~\ref{f:spec}, including generic models of Seyfert galaxies, power-law spectra and the ionizing Seyfert spectrum that includes a `big blue bump' based on the BH2013 model (Appendix C within), where we assumed a hot component (power-law) fraction of 10\% relative to the big blue bump ($k_2=k_1$ in equation 3 of BH2013).

%



The time-dependent models were run by turning on the source of ionization, waiting for the gas to reach ionization/thermal equilibrium, and then turning off the ionizing photon flux. The sound crossing time of the warm ionized layers is too long ($\gta$10 Myr) in the low density regime relevant to our study for isobaric conditions to prevail; essentially all of our results are in the isochoric limit.

We provide a synopsis of our extensive modelling in Fig.~\ref{f:AGNfading}
and Table~\ref{t:models}. 
In order to account for the Si and C ion ratios and projected column densities, we must `over-ionize' the gas, at least initially. This pushes us to a higher-impact ionization parameter at the front of the slab. Given that the fading source must also account for the \Ha\ emissivity along the Magellanic Stream, we can achieve the higher values of $u$, specifically $\log u > -3$, by considering gas at even lower density (\nH $<$ 0.01 cm$^{-3}$) consistent with the Stream's properties. Specifically, for the UV diagnostic sight lines, the \HI\ column is in the range $\log$\NH\ $=$ 17.8-18.3 when detected (Fox et al 2014, Fig. 4), although for most sight lines, only an upper limit in that range is possible. For our canonical Stream depth of $L\sim 1$ kpc, this leads to \nH\ $\sim$ 0.001 cm$^{-3}$. Such low densities lead to initially higher gas temperatures, and slower cool-down and recombination rates.



The range of densities derived in this way is reasonable.
The high end of the range explains the presence of both \HI\ and H$_2$ in absorption along the Stream (Richter et al 2013). More generally, for a spherical cloud, its mass is approximately $M_c \sim f_n\rho_c d_c^3/2$ where the subscript $n$ denotes that the filling factor refers to the neutral cloud prior to external ionization. From the projected \HI\ and \Ha\ data combined, the Magellanic Stream clouds rarely exceed $d_c \approx 300$ pc in depth and $N_c\approx 10^{21}$ cm$^{-2}$ in column, indicating total gas densities of $n_H = \rho_c/m_p \lta $ a few atoms cm$^{-3}$ in the densest regions, but extending to a low density tail (reaching to 3 dex smaller values) for most of the projected gas distribution.

In Fig.~\ref{f:AGNfading}, we see that the higher ionization parameters
(upper: $\log u=-1$, lower: $\log u=-2$) are ideal for reproducing the UV diagnostics (grey bands).
The very high photon fluxes (relative to the adopted \nH\ $\sim$ 0.001 cm$^{-3}$) generate high temperatures in the gas ($\sim 20-30,000$ K depending on $\log u$; Table~\ref{t:models}) and the harder spectrum ensures the higher ion columns (see Fig.~\ref{f:carbon}). This gas cools in time creating enhanced amounts of lower ionization states like \CII. The lower initial densities ensure the cooling time is not too rapid. UV diagnostics like \CIV/\CII\ and \SiIV/\SiII\ decline on timescales of order a few Myr. 

Note that AGN models run at higher $u$ ($\log u > -1$) are unphysical within the context of our framework. This would either require even lower gas densities in the slab, which are inconsistent with the observed column densities in the Stream, or an AGN source at Sgr A* that has super-Eddington accretion ($f_E > 1$). While such sources appear to exist around low-mass black holes (e.g. Kaaret et al 2017), we are unaware of a compelling argument for super-Eddington accretion in Seyfert nuclei (cf. Begelman \& Bland-Hawthorn 1997). In any event, going to an arbitrarily high $u$ with a hard ionizing spectrum overproduces \CV\ and higher states at the expense of \CIV.

\subsection{Constraining the lookback time of the Seyfert flash}
\label{s:UVtwo}



If Sgr A* was radiating at close to the Eddington limit within the last 0.5 Myr, the entire Magellanic Stream over the SGP would be almost fully ionized (e.g. Fig.~\ref{f:carbon}) -- this is not observed. Instead, we are witnessing the Stream at a time when the central source has switched off and the gas is cooling down. The different ions (H, \CII, \SiII, \CIV, \SiIV) recombine and cool at different rates depending on the local gas conditions. We can exploit the relative line strengths to determine a unique timescale while keeping in mind that the observed diagnostics probably arise in more than one environment.

Taken together, the \Ha\ surface brightness and the UV diagnostic ratios observed along the Stream tell a consistent story about the lookback time to the last major ionizing event from Sgr A* (cf. Figs. 6, 7, 11). These timescales are inferred from detailed modelling but the model parameters are well motivated. For a Stream distance of 75 kpc or more, the Eddington fraction is in the range $0.1 < f_E < 1$. For our model to work, we require the \Ha\ emission and UV absorption lines to arise from different regions. For the same burst luminosity, the initial ionization parameter $u_o^{\rm H\alpha}$ to account for the \Ha\ emission is $\log u_o^{\rm H\alpha} \sim -3$ impinging on gas densities above \nH\ $\sim$ 0.1 cm$^{-3}$. The initial conditions $u_o^{\rm UV}$ for the UV diagnostics are somewhat different with $\log u_o^{\rm UV} \sim -1$ to $-2$ operating with gas densities above \nH\ $\sim$ 0.001 cm$^{-3}$.

In Fig.~\ref{f:AGNfading}, the AGN models are able to account for the UV diagnostics. The time span is indicated by when both the \CIV/\CII\ and \SiIV/\SiII\ tracks fall within the grey band accommodating most of the `ionization cone' data points in Fig.~\ref{f:CIV}. The lower time limit is defined by $\log u=-2$ (both AGN model tracks in band) and the upper time limit by $\log u=-1$. Taken together, this indicates a lookback time for the AGN flash of about $2.5-4.5$ Myr. As shown in Fig. 7, the UV diagnostics are more restrictive than the \Ha\ constraint.  When looking at both figures, we must include the double-crossing time of $2 T_c \approx 0.5$ Myr (BH2013) to determine the total lookback time.


\subsection{Fading source: rapid cut-off or slow decay?}

\begin{figure}[htbp]
   \centering 
  \includegraphics[scale=0.58]{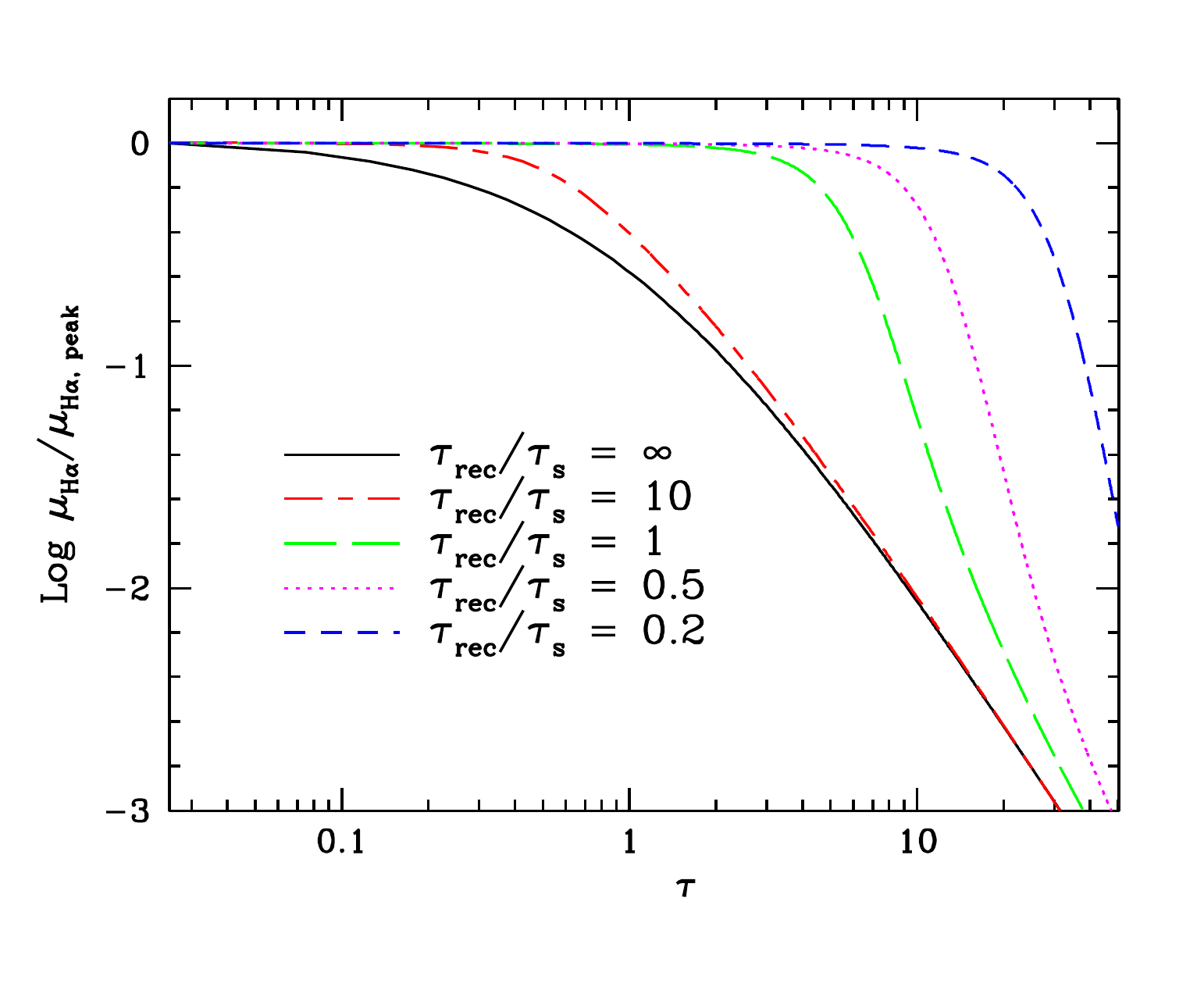}
  \caption{The predicted \Ha\ surface brightness relative to the peak value versus time $\tau$ measured in units of the recombination time. The individual curves are labeled with the ratio of the recombination time $\tau_{\rm rec}$ to the flare $e$-folding time, $\tau_s$.
 }
   \label{f:muHalpha}
   \medskip
\end{figure}

Our model assumption that the flare abruptly turned off is not necessarily correct, and the behaviour of the \Ha\ emission and the UV diagnostics can be different when the flare decay time is non-zero. To understand this behaviour, in Fig.~\ref{f:muHalpha}, we reproduce for the reader's convenience Fig. 8 from BH2013. This shows the \Ha\ surface brightness relative to the peak value as a function of $\tau$, the time since the source's flare began to decline in units of the recombination time. Each curve is labeled with the ratio of the recombination time to the $e$-folding timescale for the flare decay, $\tau_s$. Note that the limiting case $\tau_{\rm rec}/\tau_s= \infty$ is for a source that instantly turns off.

We refer the reader to Appendix A of BH2013 for mathematical details, but the important point is the following. If $\tau_{\rm rec}/\tau_s$ is small, say 0.2, the surface brightness does not begin declining until $\tau\approx 20$. This is just a reflection of the fact that if the recombination time is short compared to the source decay time, the ionization equilibrium tracks the instantaneous incident ionizing photon flux, and the flare decline takes many recombination times. If $\tau_{\rm rec}/\tau_s$ is greater than $\sim$ a few, on the other hand, then the results are nearly indistinguishable from the instant turn-off case, except for $\tau < 1$. 

\begin{figure}[htbp]
   \centering 
  \includegraphics[scale=0.5]{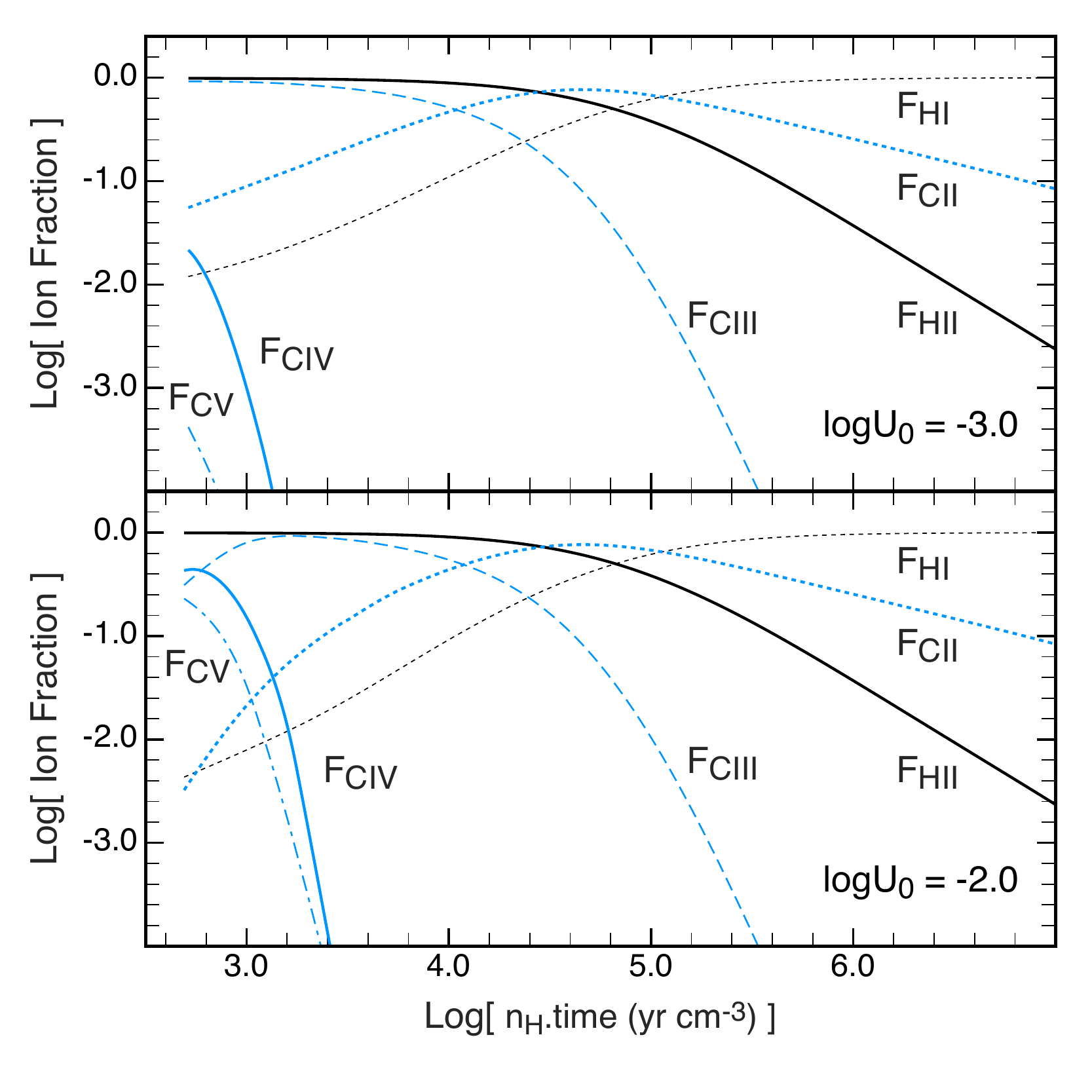}
  \caption{ {\it Mappings V} calculations for the ionization fraction (F) for different ions as a function of the product of gas density \nH\ (cm$^{-3}$) and time (years). The ionization source is our AGN power-law ($f_\nu \propto \nu^{-1}$) model (\S 3); the addition of the `big blue bump' increases the timescale by a small factor. The radiation is hitting a cold slab of gas with sub-solar metallicity ($Z=0.1 Z_\odot$). At the front face, the ionization parameter is $\log u = -3.0$ (top) and $\log u = -2.0$ (bottom). 
 }
   \label{f:ion_recomb}
   \medskip
\end{figure}

Although Fig.~\ref{f:muHalpha} shows the normalized \Ha\ surface brightness, it applies to any measure of the ionization state of the gas, in particular to \CIV/\CII. In Fig.~\ref{f:ion_recomb}, we use {\it Mappings V} to compute the time dependence of the relevant carbon ion ratios after the Stream gas has been hit by a Seyfert flare. In this model the gas has been allowed to come into photoionization equilibrium, and then the source was turned off. Results are presented for two different ionization parameters, $\log u = -2.0, -3.0$. We scale out the density dependence by using $\nH t$ as the horizontal axis.
Note that this is equivalent to plotting time $\tau$ in recombination times, as in Fig.~\ref{f:muHalpha}.

This figure illustrates two important points. First, once \CII\ becomes the dominant carbon ion, at $\log \nH t\approx 4.4$, it has a recombination time that exceeds that of \HII. Secondly, and more importantly for the Stream UV absorption line observations, \CIV\ is abundant only for a very limited range in $\log \nH t$, due to its rapid recombination. Since the UV observations show that \CIV\ and \CII\ are comparable in abundance (see Figure \ref{f:CIV}), this places a strict upper limit on the age of the burst once the gas density is known.
(Note that in the regime where the \CIV/\CII\ ratio is near the observed values, \CIII\ is the dominant carbon ion in the gas.)

For \CIV, $\tau_{\rm rec}/\tau_s$ is always much smaller (for gas of similar density) than it is for \HII ; this is why the \CIV\ abundance declines so much more rapidly compared to \HII\ in Fig.~\ref{f:ion_recomb}. It is plausible that $\tau_{\rm rec}$ for \CIV\ is short compared to $\tau_s$ (indeed, this is the likely case unless the flare decay was very abrupt or the Stream densities are unexpectedly low). Hence the carbon ionization balance will closely track the photoionization equilibrium corresponding to the instantaneous value of $\phi$ (and hence $u$), while the \Ha\ emission will reflect an earlier, larger ionizing flux.

In summary, the flare could be decaying at the present lookback time of approximately half a million years, and the carbon absorption lines (in particular, \CIV/\CII) measure the strength of the ionizing flux at that time. The brightest \Ha\ emission then reflects the peak intensity of the ionizing flux, or at least something closer to that value than what is indicated by the carbon ion ratios.

\begin{figure}[htbp]
   \centering 
  \includegraphics[scale=0.8]{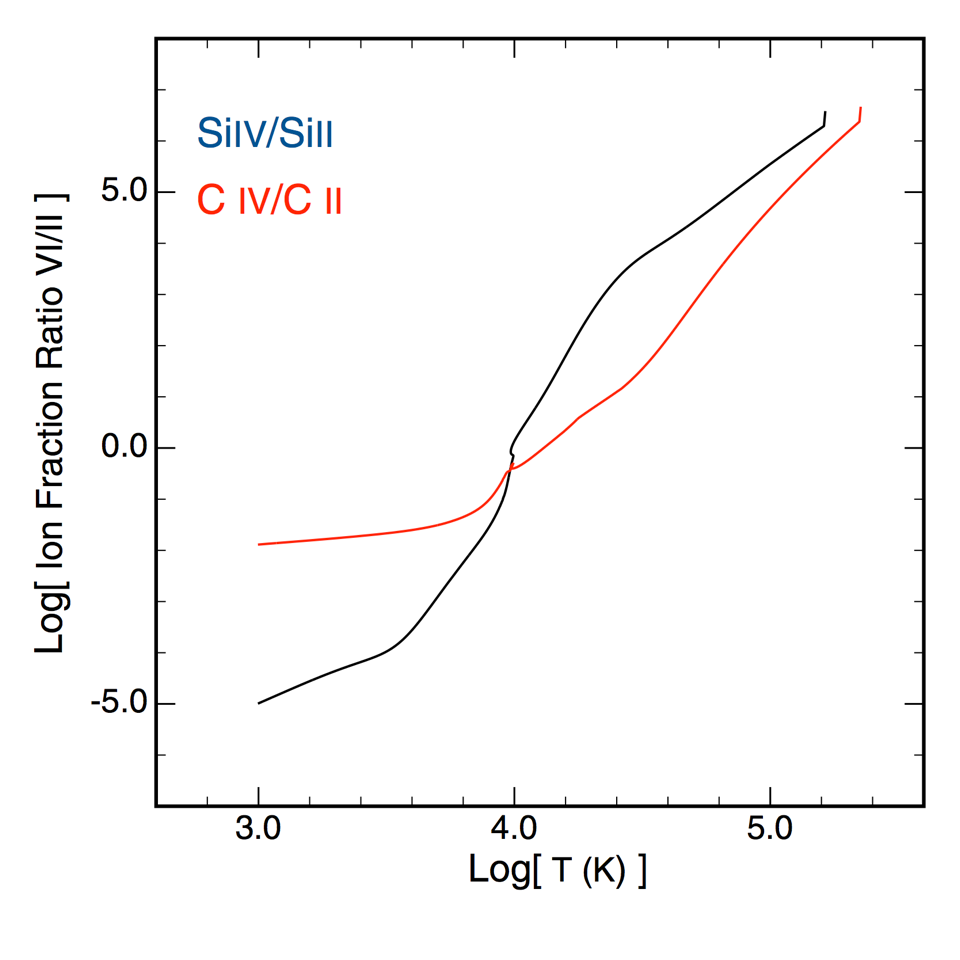}
  \caption{{\it Mappings V} calculations (assuming $Z=0.1Z_\odot$) for the ratio of two ions in a cooling gas shown for \CIV/\CII\ and \SiIV/\SiII. The tracks cross at 10$^4$ K and above 10$^5$ K relevant to photoionization and moderate shock ($v_s \sim 100$ km s$^{-1}$) zones respectively.
}
   \label{f:ion_temp}
   \medskip
\end{figure}
\subsection{Other potential sources of ionization} 

\subsubsection{Explosive shock signatures}


We find that a \textit{Fermi }bubble-like explosion in the distant past ($\sim$ 150 Myr $(v_s/500\; {\rm km\; s^{-1}})^{-1}$) $-$ moving through the Magellanic Stream today with a shock velocity $v_s$ $-$ cannot explain either the UV diagnostics or the \Ha\ emissivity, even considering the additional contribution from the photoionized precursor. The detailed modelling of Miller \& Bregman (2016) makes that very clear when extrapolated from 10 kpc (tip of the bubble) to a distance of 75 kpc or more.  The intrinsic wind velocity creating the pressure in the bubbles is of order 3000$-$10,000\kms\ (Guo \& Mathews 2012) but the wind must push aside the hot Galactic corona to reach the Magellanic Stream. 

Today, there is a strong pressure gradient across the bubbles, with a thermal pressure ($P_{\rm th}/k$) of roughly 6000 cm$^{-3}$ K at the base dropping to about 1000 cm$^{-3}$ K at the tip.  The hot shell has an outflow (shock) velocity of $v_s \approx 490$\kms\ (Mach number, ${\cal M}\approx 2-3$) pushing into an external Galactic corona with $P/k \approx 200$ cm$^{-3}$ K. If a cloud exists at the tip of the \textit{Fermi} bubbles, the combined thermal and ram-pressure shock driven into the lowest density gas drives a shock velocity of $v_s \approx 60$\kms, too weak to account for \CIV, \SiIV\ or the \Ha\ emissivity. The same holds true for the weak x-ray emission emanating from the cooling bubbles. 

In reality, the \textit{Fermi} bubbles are expected to expand and diffuse into the Galactic corona after only a few tens of kiloparsecs, such that the hot shell never reaches the Stream. To date, diffuse x-ray emission associated directly with the Magellanic Stream has never been observed and would not be expected in our scenario. 
Thus we do {\it not} believe an energetic bubble (or jet) has ever swept past the Magellanic Stream and, even if it were possible, the shock front would be too weak to leave its mark. 

For completeness, we use {\it Mappings V} to explore plausible time-dependent shock scenarios for exciting \CIV\ and \SiIV. Once again, we treat metal-poor gas and assume a 1D planar geometry at the working surface. If the shock is allowed to run indefinitely, it cools down to a mostly neutral phase near 100~K. Under these conditions, \CII\ and \SiII\ ionization fractions steadily rise with respect to the higher ionization states. If we truncate the cooling shock at 10$^4$ K, \CIV/\CII\ and \SiIV/\SiII\ are both less than 0.1 for fast shocks ($v_s \gta 100$ km s$^{-1}$), but diverge for slow shocks, e.g. $v_s = 60$ km s$^{-1}$ gives \CIV/\CII\ $\approx$ 0.01, and \SiIV/\SiII\ $\approx$ 25. These are manifestly inconsistent with observations.

In Fig.~\ref{f:ion_temp}, we compute the \CIV/\CII\ and \SiIV/\SiII\ ion ratios vs. the ionized gas temperature, $T_e$. This ratio is {\it not} independent of the gas abundances for metal-poor gas; the calculation is undertaken with $Z = 0.1 Z_\odot$. The ion ratios reach parity at $T_e \approx 10^4$ K and for $T_e > 10^{5.3}$ K. Taken together, the ion ratios are certainly consistent with photoionization but their convergence at higher temperature suggests another possible origin. The \CIV/\CII\ ratio, like the \SiIV/\SiII\ ratio (both with up to 0.5 dex of scatter), is of order unity and is enhanced in a region over the South and North Galactic Poles (see Fig.~\ref{f:CIV}). So are there other ways to increase the 
gas temperature without photoionization or shocks from blast waves? We address this issue in the next section.

\subsubsection{Shock cascade and turbulent mixing}

Bland-Hawthorn et al (2007) and Tepper-Garcia et al (2015) consider the case of the Magellanic \HI\ stream being ablated by the diffuse hot halo. They show that the post-shock cooling gas ($v_s < 20$ km s$^{-1}$) in a `shock cascade' is generally too weak along the Magellanic Stream to power the \Ha\ emission, particularly at the newly established distance of $D > 75$ kpc (cf. Barger et al 2017). The post-shock temperature ($<10^4$K) is too low to produce high-ionization species, even in the high-energy tail of the particle distribution (cf. Fig.~\ref{f:ion_temp}). But a shock cascade can still be important even if it does not account for the observed spectral signatures directly. For example, it can help to break down the cold gas and enable interchange with the hot halo.

A major uncertainty along the Stream is the degree of mixing between the cold clouds and the hot coronal gas; a shearing boundary layer can give rise to intermediate gas phases with a mean temperature of order $\sqrt{T_{\rm hot} T_{\rm cold}}$ and therefore a broad range of ionization states (Ji et al 2019; Begelman \& Fabian 1990). This process is driven by either Kelvin-Helmholz (KH) instabilities at the hot/cold interface, or turbulence in the hot corona for which there are few constraints presently. The outcome depends on the fraction of mass of hot gas deposited into the mixing layer, and the efficiency of hydrodynamic mixing.

To our knowledge, there have only been two hydrodynamic studies of this turbulent regime that incorporate consistent non-equilibrium ionization, i.e. Esquivel et al (2006; MHD) and Kwak \& Shelton (2010; HD). Notably, Kwak \& Shelton (2010) find, much like for conductive interfaces (see below), that the low and high ionization states arise from very low column gas ($\lesssim 10^{13}$ cm$^{-2}$). While mixing in sheared layers surely exists at the contact surface of the \textit{Fermi} bubbles (Gronke \& Oh 2018; Cooper et al 2008), it is unclear if these processes are possible at the Stream's distance over the South Galactic Pole where the coronal density is low ($\sim$ a few $\times$ $10^{-5}$ cm$^{-3}$).

Several authors have discussed the idea of conductive interfaces in which cool/warm clouds evaporate and hot gas condenses at a common surface where colliding electrons transport heat across a boundary (Gnat, Sternberg \& McKee 2010; Armillotta et al 2017). The gas tends to be `under-ionized' compared to gas in ionization equilibrium which enhances cooling in the different ions. But Gnat et  al (2010) show that the non-equilibrium columns are always small ($\lesssim 10^{13}$ cm$^{-2}$) and an order of magnitude below the median columns detected by Fox et al (2014).




For full consistency, the shock cascade model is 
an appropriate framework for a mixing layer calculation but a self-consistent radiative MHD code to achieve this has yet to be developed. Our first models predict
projected line broadening up to $\sigma \approx 20$\kms\ in \HI\ or warm ion transitions (Bland-Hawthorn et al 2007). It is possible that running models with intrinsically higher resolution, one can broaden the absorption line kinematics and increase the column densities further through line-of-sight projections.
An important future constraint is to map the relative distributions of warm ionized, warm neutral and cold neutral hydrogen gas at high spectral/spatial resolution along the Stream.

Presently, we do not find a compelling case for dominant processes beyond static photoionization from a distant source. All of these processes may have more relevance to the \textit{Fermi} bubbles and to high velocity clouds (HVCs) much lower in the Galactic halo ($D \ll 75$ kpc). For the HVCs, such arguments have been made (Fox et al 2005). Before an attempt is made to understand the Stream in this context, it will be crucial to first demonstrate how turbulent mixing has contributed to UV diagnostics observed towards low-latitude clouds.

\subsection{Correlations between the observed diagnostics along the Magellanic Stream}

\subsubsection{The scatter in the \Ha\ emission relative to \HI}
\label{s:patchy}

Ideally, we would be able to bring together all spectroscopic information within a cohesive framework for the Magellanic Stream in terms of its origin, internal structure, ionization and long-term evolution (e.g. Esquivel et al 2006; Tepper-Garcia et al 2015). As implied in the last section, various parts of the problem have been tackled in isolation, but an overarching scheme covering all key elements does not exist today. For such a complex interaction, we must continue to gather rich data sets across the full electromagnetic spectrum (Fox et al 2019). Our work has concentrated on both absorption and emission lines observed with very different techniques, effective beam sizes and sensitivities. We now consider what one might learn in future when both absorption and emission measures have comparable sensitivities and angular resolution. This may be possible in the era of ELTs, at least for the \Ha-bright regions.

Figure 12 of Barger et al (2017) shows the lack of any correlation between the \Ha\ detections and the projected \HI\ column density. The emission measures mostly vary over about a factor of five, from $\sim 30$ to 160 mR; there are two exceptionally bright knots along the Stream with $400 \lesssim \mu(\Ha) \lesssim 600$ mR. The total H column (\HI\ + \HII) today is high enough to absorb a significant fraction of incident UV photons across much of the Stream {\it if the Sgr A* source currently radiates far below the Eddington limit}. This simple observation is consistent with the nuclear flare having shut down and the Stream's recombination emission fading at a rate that depends only on the local gas density. For completeness, we mention one more possibility which is somewhat fine-tuned and therefore less plausible. It is possible that at the lookback time ($2T_c\approx 0.5$ Myr) at which we observe the Stream emission (for a distance of 75 kpc), the Galaxy's nuclear emission is still far above the present-day value and the spread in emission measures is dominated by column density variations along the lines of sight.

Assume for a moment that variations in $N/N_{\rm cr}$ are unimportant. In principle, the power spectrum of the \Ha\ patchiness constrains both the gas densities and the time since the radiation field switched off, since the scatter increases with the passage of time (up until the recombination time for the lowest density gas is reached), due to the spread in $\tau_{\rm rec}$; see Figure 4 in BH2013. However, there are several complications. The predicted range of $\mu_{{\rm H}\alpha}$ as a function of time depends on the distribution of gas densities within the Stream. At present, however, the observable range in \Ha\ surface brightness is limited by the moderate S/N of most of the detections. 

An additional complication is that, at fixed density \nH, lines of sight with $N < \Ncr$ will be fainter in \Ha\ by the ratio $N/\Ncr$, as discussed above. Finally, the observed patchiness is likely to be heavily filtered by the angular resolution of the observer's beam (Tepper-Garcia et al 2015). In future, it may be possible to sort out these issues with knowledge of the {\it total} hydrogen column density along the Stream from independent sources of information, e.g., soft x-ray shadowing by the Stream projected against the cosmic x-ray background (e.g. Wang \& Taisheng 1996).

\subsubsection{The scatter in the UV absorption lines relative to \HI}
\label{s:UVone}

For absorption lines, it is the column density $N_p$ that matters, not the product $n_e N_p$. In other words, the \Ha\ emission from low-density regions with large columns is, in effect, being scaled down by their low densities, but this is not true for the UV absorption lines. So, in this model, the prominence of the lowest-density regions in the absorption-line observations will be even more pronounced than it is for the \Ha\ emission: they not only stay more highly ionized for longer, because of the longer recombination times, but they also arise in the largest H column densities (Fig.~\ref{f:carbon}), and that is what the absorption-line diagnostics are sensitive to.

What this argument does not determine is whether the carbon ionization state (as measured by the \CIV/\CII\ ratio) resembles what we are seeing for some reasonable period of time after the source turns off, or whether the only applicable models are ones in which the ionization state has not really had time to change. That still favors the lowest-density regions, however, for the reasons just outlined.

In general, for the assumed tubular geometry of the Magellanic Stream, we expect higher densities to roughly correspond to larger column densities. However, in the flare ionization model, as noted above, the densest regions recombine the fastest, and thus fade quickly in \Ha\ and lose their \CIV\ rapidly once the flare has switched off. In the flare model, as long as the gas column densities along the Stream are greater than the critical column needed to soak up all of the ionizing photons, the density/column density anticorrelation (lower density regions have larger ionized columns) is baked in by the physics, and so in this case we anticipate a positive correlation between the \Ha\ emission and the \CIV\ absorption strength.

There are two caveats: the correlation (1) only arises if the low-density regions still have significant \CIV\ fractions (i.e., they have not had time to recombine to low ionization states); (2) would not hold if the \CIV\ is coming mostly from regions where the density is so low that the total column is lower than the critical column, i.e., density-bounded rather than radiation-bounded sightlines. In the latter case, the \Ha\ emission will also be weaker than our model predicts, by the ratio of the actual column to the critical column.  The \HI/\Ha\ comparison above was possible because of the comparable ($0.1-1$ degree) beam size for both sets of observations. Unfortunately, the UV absorption lines have an effective beam size that is orders of magnitude smaller then either the optical or radio detections. An additional problem are the short timescales associated with \CIV\ recombination relative to \Ha\ and \CII\ as we discuss below.


\section{Discussion}

There is nothing new about the realisation of powerful episodic behaviour erupting from the nuclei of disk galaxies (q.v. Mundell et al 2009). Some of these events could be close analogues to what we observe today in the Milky Way (cf. NGC 3079: Li et al 2019; Sebastian et al 2019).
Since 2003, many papers present evidence for a powerful Galactic Centre explosion from radio, mid-infrared, UV, x-ray and $\gamma$ ray emission. The remarkable discovery of the $\gamma$-ray
bubbles (Su et al 2010) emphasized the extraordinary power of the event. The dynamical ($2-8$ Myr) and radiative ($2.5-4.5$ Myr) timescales overlap, with possible evidence
that the jet/wind break-out (Miller \& Bregman 2016) preceded the radiative event (this work; BH2013).
Conceivably, if the error estimates are reliable, this time difference is real, i.e. the explosive event was needed to clear a path for the ionizing radiation.

In the search for a singular event that
may have triggered Sgr A* to undergo a Seyfert phase,
we find the link to the central star streams and young clusters made
by Zubovas \& King (2012) to be compelling.
Against a backdrop of ancient stars, Paumard et al
(2006) review the evidence for a young stellar ring with well constrained ages of $4-6$ Myr. The same connection may extend to the circumnuclear star clusters that fall within the same age range (Simpson 2018). 
Intriguingly, Koposov et al (2019) have recently discovered a star travelling at 1750\kms\ that was ejected from the Galactic Centre some 4.8 Myr ago. It is tempting to suggest this was also somehow connected with the major gas accretion event at that time, i.e. through stars close to the black hole being dislodged.

This could reasonably be made to fit with the shorter timescale
($T_o=3.5\pm 1$ Myr) for the flare if the event was sufficiently cataclysmic in the vicinity of Sgr A* to directly fuel the inner
accretion disk. Accretion timescales of infalling
gas being converted to radiative output can be as
short as 0.1$-$1 Myr (Novak et al 2011) 
although Hopkins et al (2016) argue for
a longer viscosity timescale. We now consider how the field can advance in future years with sufficient observational resources.

\smallskip\noindent{\sl Towards a complete 3D map of halo clouds.} The most successful approach for absorption line 
detections along the Magellanic Stream has been to target UV-bright ($B < 14.5$)
background AGN and quasars (Fox et al 2013, 2014). 
In future, all-sky high-precision photometric imaging (e.g. LSST) will allow us to easily 
identify a population of UV-bright, metal-poor halo stars 
with well established photometric distances.
Targetting some stars ahead and behind the Stream will improve distance brackets for the
Stream and provide more information on the nature 
of the recent Seyfert outburst. There are many potential targets across the sky.
The {\sl Galaxia} model of the Galaxy (Sharma et al 2011)
indicates there is one metal poor giant per square degree
brighter than $B=14.5$
in the Galactic halo out to the distance of the Stream,
with a factor of six more at $B=16$ which can be
exploited in an era of ELTs.
In principle, it will be possible to determine good distances to all neutral and ionized HVCs from distance
bracketing across the entire halo, particularly within 50 kpc or so. 

The high-velocity \HI\ clouds lie almost
exclusively close to the Galactic Plane, i.e.  outside the \HI-free cones identified
by Lockman \& McClure-Griffiths (2016). There are highly ionized HVCs seen all over the sky found in \OVI\ absorption but not in \HI\ emission (Sembach et al 2003). The \OVI\ sky covering fraction is in the range 60-80\% compared to the \HI\ covering 
fraction at about 40\%. The use of near-field clouds to trace the ionization cones is hampered by the presence of ionized gas entrained by the x-ray/$\gamma$ ray bubbles
(Fox et al 2015; Bordoloi et al 2017; Savage et al 2017; Karim et al 2018). But we anticipate that the ionization cones (Fig.~\ref{f:nidever}) and the \textit{Fermi} bubbles (Fig.~\ref{f:fermi}) are filled with hundreds of distinct, fully ionized HVCs.

\smallskip\noindent{\sl Magellanic Screen - viewing the AGN along many sight lines.}
The Magellanic Stream provides us with a 
fortuitous absorber for intersecting ionizing radiation 
escaping from the Galactic Centre. This `Magellanic Screen'
extends over 11,000 square degrees (Fox et al 2014) and enables us
to  probe the complexity of the emitter over
wide solid angles. Our simple adoption of the Madau
model predicts a centrosymmetric pattern along some
arbitrary axis.
But many models produce
anisotropic radiation fields, e.g. jets (Wilson
\& Tsvetanov 1994), thick 
accretion disks (Madau 1988), warped accretion
disks (Phinney 1989; Pringle 1996), dusty
tori (Krolik \& Begelman 1988; Nenkova et al 2008)
binary black hole. More measurements along the
Stream may ultimately shed more light on the
recent outburst from Sgr A* and its immediate
surrounds. The strongest constraint comes from
variations in the ionization parameter $u$
(Tarter et al 1969), but detecting
second order effects from the spectral slope may be possible (e.g. Acosta-Pulido et al 1990), although time-dependent ionization complicates matters (\S~\ref{s:agn}).

This suggests a future experiment.
Consider an ionization pattern defined by an axis tilted 
with respect to 
the Galactic poles. Here we are assuming something like
the \CIV/\CII\ line ratio to measure spectral `hardness' {\cal H}
or ionization parameter $u$ over the sky. We can now fit 
spherical harmonics to the all-sky distribution to establish
the dominant axis of a centrosymmetric pattern (e.g. Fixsen
et al 1994). For
illustration, we
project our crude fit in Fig.~\ref{f:CIV} as a sine wave
in Magellanic longitude. To be useful, we need
many more sight lines over the sky.

We are far from a convincing narrative for Sgr A* as we are for any supermassive black hole. These fascinating sources are seeded and grow rapidly in the early universe, and then accrete more slowly with the galaxy's evolution over billions of years. Just how they interact and influence that evolution is an outstanding problem in astrophysics.
We live in hope that this new work may
encourage accretion disk modellers (e.g. GR-R-MHD codes; McKinney et al 2014) to consider the UV outburst
in more detail, and to predict the emergent 
radiation and timescale to aid future comparisons
with observations. Ultimately, such models will need to be integrated into fully cosmological models of galaxy formation and evolution.

\section{Acknowledgment}
JBH is supported by a Laureate Fellowship from the Australian Research Council. JBH also acknowledges a 2018 Miller Professorship at UC Berkeley where the first draft was completed. MG acknowledges funding from the University of Sydney. WHL is supported by China-Australia Scholarship funds for short-term internships. We are particularly grateful to James Josephides (Swinburne University) and Ingrid McCarthy (ANU) for the movie rendition of the Magellanic Stream being ionized by the accretion disk around Sgr A* (see the caption to Fig. 4). Over the past few years, we have benefitted from key insights and suggestions: we thank Chris McKee, Luis Ho, Jenny Greene, Will Lucas, Carole Mundell, Dipanjan Mukherjee, Roger Blandford, Lars Hernquist, Jerry Ostriker, Ramesh Narayan and an anonymous referee, assuming of course they are not in the list already.

\appendix

\section{A. Emission measures}
In order to compare our model with the H$\alpha$ observations,
we adopt physically motivated units that relate the ionizing photon flux at a distant cloud to the resultant H$\alpha$ emission.  It is convenient to relate the plasma column emission rate to a photon surface brightness. Astronomical research on diffuse
emission (e.g. WHAM survey $-$ Reynolds et al 1998) use the
Rayleigh unit introduced by atmospheric 
physicists (q.v. Baker \& Romick 1976)
which is a unique measure of {\it photon} intensity; 1 milliRayleigh
(mR) is equivalent to $10^3/4\pi$ photons cm$^{-2}$ s$^{-1}$
sr$^{-1}$.  The emission measure ${\cal E}_m$ for a plasma with
electron density $n_e$ is given by (e.g. Spitzer 1978)
\bee
{\cal E}_m = \int f_i n_e^2\ {\rm d}z \ \ \ \ {\rm cm^{-6} \ pc}
\label{e:em}
\eee
which is an integral of H recombinations along the line of sight $z$
multiplied by a filling factor $f_i$. The suffix $i$ indicates that we
are referring to the volume over which the gas is ionized. For a
plasma at 10$^4$K, ${\cal E}_m(\rm H\alpha)=1$ cm$^{-6}$ pc is
equivalent to an H$\alpha$ surface brightness of 330 milliRayleighs
(mR).  In cgs units, this is equivalent to $1.9\times 10^{-18}$ erg
cm$^{-2}$ s$^{-1}$ arcsec$^{-2}$ which is a faint spectral
feature in a 1 hr integration using a slit spectrograph on an 8m
telescope. But for the Fabry-Perot `staring' technique employed in
Fig.~\ref{f:WHAM}, this is an easy detection if the diffuse emission
uniformly fills the aperture. We refer to the Stream H$\alpha$
emission as relatively bright because it is much brighter than
expected for an optically thick cloud at a distance of 55 kpc or 
more from the Galactic Centre.


\section{B. Photoionization due to a Seyfert flare event}

{\sl Mappings V} has been adapted to incorporate the time-dependent calculations in BH2013. Here a gas slab is ionized by a burst of radiation, which is then allowed to cool down over millions of years. These calculations, which use a wide range of ionizing sources (see Fig.~\ref{f:spec}), are specifically aimed at C ions (Table 2) and Si ions (Table 3) observed at UV wavelengths (e.g. Fox et al 2014). A summary of the initial model parameters is given in Table 1.

\begin{table}[!ht]
\caption{Photoionization models: initial conditions in the gas slab when the ionizing source is switched on. The columns are (1) ionizing source in Fig.~\ref{f:spec}, (2) ionization parameter, (3) initial electron temperature, and remaining columns give predicted column densities for C ions as indicated.}
\begin{center}
\begin{tabular}{cccccccccc}
\hline
Model&$\log u$&T$_e$&N(CVII)&N(CVI)&N(CV)&N(CIV)&N(CIII)&N(CII)&\\
\hline
\hline
BLS (a) &-1.0&2.99E+04&1.82E+12&8.32E+13&6.86E+14&5.23E+13&5.21E+12&9.79E+09& \\
&-2.0&2.13E+04&4.50E+09&2.55E+12&2.64E+14&3.21E+14&2.39E+14&2.57E+12& \\
&-3.0&1.57E+04&1.43E+05&1.14E+09&1.40E+12&4.74E+13&7.09E+14&7.10E+13& \\
BLS + soft Compton (a) &-1.0&3.50E+04&1.29E+13&2.23E+14&5.75E+14&1.68E+13&1.38E+12&3.81E+09& \\
&-2.0&2.31E+04&5.99E+10&1.35E+13&4.60E+14&2.24E+14&1.30E+14&1.70E+12& \\
&-3.0&1.63E+04&3.85E+06&1.06E+10&4.46E+12&6.31E+13&6.86E+14&7.50E+13& \\
\hline
NLS (b) &-1.0&3.26E+04&1.49E+12&8.09E+13&7.29E+14&1.66E+13&9.14E+11&1.99E+09& \\
&-2.0&2.26E+04&6.72E+09&4.63E+12&5.34E+14&2.06E+14&8.38E+13&9.53E+11& \\
&-3.0&1.61E+04&3.72E+05&3.12E+09&4.42E+12&7.94E+13&6.77E+14&6.74E+13& \\
NLS + soft Compton (b)  &-1.0&3.67E+04&1.84E+13&2.68E+14&5.32E+14&9.82E+12&6.28E+11&1.92E+09& \\
&-2.0&2.35E+04&1.06E+11&2.05E+13&5.49E+14&1.78E+14&8.00E+13&1.08E+12& \\
&-3.0&1.63E+04&8.14E+06&1.96E+10&6.55E+12&7.58E+13&6.71E+14&7.51E+13& \\
\hline
BH2013 (c) &-1.0&3.18E+04&6.02E+11&5.03E+13&7.48E+14&2.76E+13&2.24E+12&5.16E+09& \\
&-2.0&2.24E+04&2.15E+09&2.25E+12&4.23E+14&2.51E+14&1.51E+14&1.88E+12& \\
&-3.0&1.61E+04&7.24E+04&1.12E+09&2.57E+12&6.14E+13&6.90E+14&7.50E+13& \\
\hline
Star cluster (d) &-1.0&1.55E+04&\ldots&\ldots&1.07E+10&2.83E+14&5.45E+14&1.24E+12& \\
&-2.0&1.50E+04&\ldots&\ldots&1.21E+08&3.81E+13&7.73E+14&1.78E+13& \\
&-3.0&1.36E+04&\ldots&\ldots&2.67E+05&1.83E+12&6.53E+14&1.74E+14& \\
\hline
PL, $\alpha=-1.0$ (e) &-1.0&4.11E+04&1.34E+14&4.40E+14&2.49E+14&5.41E+12&5.41E+11&2.48E+09& \\
&-2.0&2.43E+04&1.28E+12&5.93E+13&4.77E+14&1.73E+14&1.17E+14&2.08E+12& \\
&-3.0&1.65E+04&1.71E+08&9.92E+10&1.01E+13&5.94E+13&6.67E+14&9.20E+13& \\
PL, $\alpha=-1.5$ (e) &-1.0&3.45E+04&1.30E+13&2.19E+14&5.68E+14&2.57E+13&3.51E+12&1.32E+10& \\
&-2.0&2.31E+04&4.74E+10&1.03E+13&3.47E+14&2.37E+14&2.30E+14&4.20E+12& \\
&-3.0&1.62E+04&3.45E+06&9.12E+09&3.81E+12&4.41E+13&6.79E+14&1.02E+14& \\
PL, $\alpha=-2.0$ (e) &-1.0&3.025E+04&5.902E+11&5.411E+13&6.936E+14&6.714E+13&1.349E+13&4.77E+10& \\
&-2.0&2.167E+04&9.969E+08&1.126E+12&1.794E+14&2.555E+14&3.853E+14&7.76E+12& \\
&-3.0&1.552E+04&8.440E+03&5.755E+08&1.109E+12&2.762E+13&6.764E+14&1.24E+14& \\
\hline
ULX100 (f)  &-1.0&3.70E+04&6.49E+14&1.70E+14&9.38E+12&1.06E+11&8.14E+09&7.42E+07& \\
&-2.0&2.07E+04&8.62E+13&3.29E+14&2.67E+14&8.64E+13&5.87E+13&2.01E+12& \\
&-3.0&1.43E+04&1.20E+10&5.05E+11&4.60E+12&8.41E+12&5.75E+14&2.34E+14& \\
ULX1000 (f) &-1.0&4.00E+04&6.40E+14&1.80E+14&9.80E+12&9.01E+10&6.20E+09&5.12E+07& \\
&-2.0&2.19E+04&8.52E+13&3.52E+14&2.88E+14&6.73E+13&3.51E+13&1.02E+12& \\
&-3.0&1.50E+04&1.71E+10&8.43E+11&8.30E+12&1.61E+13&6.12E+14&1.89E+14& \\
\hline
40,000 K Star (g) &-1.0&1.527E+04&\ldots&\ldots&2.391E+03&5.526E+12&8.217E+14&1.80E+12& \\
&-2.0&1.489E+04&\ldots&\ldots&\ldots&5.158E+11&8.104E+14&1.81E+13& \\
&-3.0&1.356E+04&\ldots&\ldots&\ldots&2.403E+10&6.584E+14&1.70E+14& \\
\hline
\end{tabular}
\end{center}
\label{t:carbon}
\end{table}%

\begin{table}[!ht]
\caption{Photoionization models: initial conditions in the gas slab when the ionizing source is switched on. The columns are (1) ionizing source in Fig.~\ref{f:spec}, (2) ionization parameter, (3) initial electron temperature, and remaining columns give predicted column densities for Si ions as indicated.}
\begin{center}
\begin{tabular}{cccccccccc}
\hline
Model&$\log u$&T$_e$&N(SiVII)&N(SiVI)&N(SiV)&N(SiIV)&N(SiIII)&N(SiII)&\\
\hline
\hline

BLS (a) &-1.0&2.99E+04&3.61E+13&4.36E+13&1.34E+13&3.29E+11&5.92E+10&7.36E+08\\
&-2.0&2.13E+04&7.47E+11&1.09E+13&4.33E+13&2.27E+13&2.12E+13&7.84E+11\\
&-3.0&1.57E+04&9.66E+07&1.36E+10&6.25E+11&1.23E+13&7.02E+13&1.65E+13\\
BLS + soft Compton (a)  &-1.0&3.50E+04&4.85E+13&1.09E+13&5.05E+11&3.96E+09&4.92E+08&6.72E+06\\
&-2.0&2.31E+04&1.60E+13&4.59E+13&2.90E+13&4.73E+12&2.90E+12&1.10E+11\\
&-3.0&1.63E+04&1.14E+10&4.03E+11&3.30E+12&2.00E+13&6.14E+13&1.45E+13\\
\hline
NLS (b) &-1.0&3.26E+04&4.22E+13&4.17E+13&7.82E+12&9.68E+10&1.26E+10&1.67E+08\\
&-2.0&2.26E+04&1.69E+12&2.08E+13&5.13E+13&1.52E+13&1.03E+13&4.00E+11\\
&-3.0&1.61E+04&2.85E+08&4.08E+10&1.28E+12&1.68E+13&6.33E+13&1.81E+13\\
NLS + soft Compton (b)  &-1.0&3.67E+04&4.20E+13&6.25E+12&1.85E+11&9.93E+08&9.68E+07&0\\
&-2.0&2.35E+04&2.52E+13&4.86E+13&1.99E+13&2.57E+12&1.20E+12&4.51E+10\\
&-3.0&1.63E+04&3.26E+10&7.93E+11&4.28E+12&2.46E+13&5.50E+13&1.49E+13\\
\hline
BH2013 (c) &-1.0&3.18E+04&3.21E+13&5.10E+13&1.30E+13&2.15E+11&3.54E+10&4.70E+08\\
&-2.0&2.24E+04&8.02E+11&1.56E+13&5.16E+13&1.66E+13&1.45E+13&5.52E+11\\
&-3.0&1.61E+04&1.47E+08&3.07E+10&1.28E+12&1.43E+13&6.84E+13&1.55E+13\\
\hline
Star cluster (d) &-1.0&1.55E+04&\ldots&\ldots&3.86E+13&4.17E+13&1.93E+13&9.85E+10\\
&-2.0&1.50E+04&\ldots&\ldots&1.03E+12&1.79E+13&7.82E+13&2.60E+12\\
&-3.0&1.36E+04&\ldots&\ldots&1.86E+09&1.83E+12&8.14E+13&1.64E+13\\
\hline
PL, $\alpha=-1.0$ (e) &-1.0&4.11E+04&2.24E+13&1.74E+12&3.12E+10&1.30E+08&1.44E+07&\ldots\\
&-2.0&2.43E+04&3.92E+13&4.07E+13&1.07E+13&9.84E+11&4.99E+11&1.90E+10\\
&-3.0&1.65E+04&1.43E+11&1.87E+12&6.51E+12&2.45E+13&5.50E+13&1.17E+13\\
PL, $\alpha=-1.5$ (e) &-1.0&3.45E+04&4.95E+13&1.28E+13&7.14E+11&7.07E+09&1.23E+09&1.82E+07\\
&-2.0&2.31E+04&1.33E+13&4.34E+13&3.26E+13&4.98E+12&4.38E+12&1.68E+11\\
&-3.0&1.62E+04&9.49E+09&3.82E+11&3.74E+12&1.60E+13&6.75E+13&1.20E+13\\
PL, $\alpha=-2.0$ (e)  &-1.0&3.03E+04&4.37E+13&4.04E+13&7.57E+12&1.78E+11&4.94E+10&7.01E+08\\
&-2.0&2.17E+04&1.69E+12&1.92E+13&4.60E+13&1.30E+13&1.91E+13&7.23E+11\\
&-3.0&1.55E+04&3.87E+08&5.34E+10&1.65E+12&9.78E+12&7.52E+13&1.30E+13\\
\hline
ULX100 (f) &-1.0&3.70E+04&4.33E+11&6.05E+09&2.50E+07&\ldots&\ldots&\ldots\\
&-2.0&2.07E+04&3.81E+13&7.52E+12&4.78E+11&1.03E+11&1.47E+10&9.11E+08\\
&-3.0&1.43E+04&5.56E+11&1.24E+12&7.57E+11&2.22E+13&1.24E+13&6.19E+13\\
ULX1000 (f)  &-1.0&4.00E+04&4.31E+11&5.33E+09&1.87E+07&\ldots&\ldots&\ldots\\
&-2.0&2.19E+04&3.68E+13&6.45E+12&3.51E+11&4.21E+10&5.86E+09&2.64E+08\\
&-3.0&1.50E+04&1.30E+12&2.78E+12&1.88E+12&3.06E+13&1.70E+13&4.55E+13\\
\hline
40,000 K Star (g) Star&-1.0&1.53E+04&1.00E+00&\ldots&2.45E+12&6.51E+13&3.20E+13&1.67E+11\\
&-2.0&1.49E+04&1.00E+00&\ldots&4.00E+10&1.70E+13&8.00E+13&2.70E+12\\
&-3.0&1.36E+04&9.95E-01&\ldots&7.11E+07&1.70E+12&8.15E+13&1.64E+13\\
\hline
\end{tabular}
\end{center}
\label{default}
\end{table}


\end{document}